\documentclass{article}

\usepackage{arxiv}

\usepackage[utf8]{inputenc} 
\usepackage[T1]{fontenc}    
\usepackage{hyperref}       
\usepackage{url}            
\usepackage{booktabs}       
\usepackage{amsfonts}       
\usepackage{nicefrac}       
\usepackage{microtype}      
\usepackage{lipsum}
\usepackage{graphicx}

\usepackage{pdfpages}

\newcommand{\Ot}{\hat{O^t}}

\title{Going beyond communication intensity for estimating tie strengths in social networks}

\author{
  Javier ~Ureña-Carrion \\
  Department of Computer Science\\
  Aalto University\\
  Espoo, Finland \\
  \texttt{javier.urenacarrion@aalti.fi} \\
   \And
 Jari ~Saramäki \\
  Department of Computer Science\\
  Aalto University\\
  Espoo, Finland \\
    \And
  Mikko ~Kivelä \\
Department of Computer Science\\
  Aalto University\\
  Espoo, Finland \\
}

\begin{document}
\maketitle

\begin{abstract}
Even though the concept of tie strength is central in social network analysis, it is difficult to quantify how strong social ties are. One typical way of estimating tie strength in data-driven studies has been to simply count the total number or duration of contacts between two people. This, however, disregards many features that can be extracted from the rich data sets used for social network reconstruction. Here, we focus on contact data with temporal information. We systematically study how features of the contact time series are related to topological features usually associated with tie strength. We analyze a large mobile-phone dataset and measure a number of properties of the call time series for each tie, and use these to predict the so-called neighbourhood overlap, a feature related to strong ties in the sociological literature. We observe a strong relationship between temporal features and the neighbourhood overlap, with many features outperforming simple  contact counts. Features that stand out include the number of days with calls, number of bursty cascades, typical times of contacts, and temporal stability. Our results suggest that these measures could be adapted for use in social network construction and indicate that the best results can be achieved by combining multiple temporal features. 
\end{abstract}


\section*{Introduction}


During the past few decades, the use of auto-recorded data, such as mobile phone logs or data from online platforms, has expanded our understanding of human dynamics and networks \cite{Onnela2007, Candia2008, Saramaki2015, Borgatti2009}. Such data have also been useful in applications ranging from spreading dynamics \cite{KarsaiSmall} to human mobility \cite{Gonzalez2008}, recovery in disaster areas \cite{Li2019}, and health-care optimization \cite{Altuncu2019}. In particular in social network studies, the \textit{strength of a tie} is a central concept associated with the qualitative value that people place on relationships. Tie strength is not, however, something that can be directly measured or quantified \cite{Carmines1982, Marsden1984, Marsden2012}. Therefore, one has to rely on proxies.  For networks reconstructed from data on communication events, such as call networks, a common approach is to use a measure of \emph{communication intensity} as a proxy \cite{Onnela2007, Karsai2014, Krings2012, Saramaki2015, Kovanen2013, Blondel2015, Park2018, TimeAllocation, Saramaki2014}. Communication intensity can be defined as the total number of communication events or the total time spent communicating across a tie. 
One motivation behind this choice is that intense communication implies temporal and sometimes even financial commitment to a relationship \cite{Onnela2007}. Communication intensity is, however, an aggregated measure that discards a lot of possibly relevant information contained in the underlying time series of dyadic interactions. This temporal information is at the focus of the present paper.

In addition to such internal details of a social tie, the network structure surrounding a tie is known to be informative about the nature of the tie. The seminal paper "The strength of weak ties" \cite{Granovetter} by Mark Granovetter was one of the first efforts to relate tie strength with local network structure and to connect the micro and macro levels by considering the role of weak ties in diffusion and social mobility. Granovetter argued that strong ties tend to be associated with overlapping circles of friendship, while 
 weak ties serve as bridges between such circles. This implies that weak ties serve are more important for network-wide information diffusion than strong ties. 

\begin{figure}
    \centering
    \includegraphics[width=.9\textwidth]{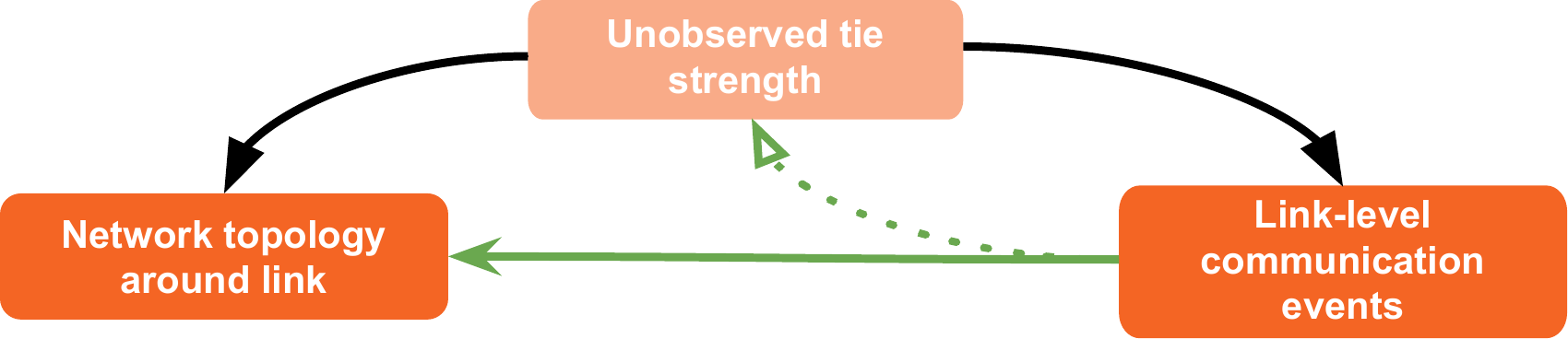}
    \caption{Representation of our conceptualization of tie strength as a latent variable that drives both network topology and patterns of human communication (black arrows). While the tie strength is unobserved, we argue that using characteristic temporal features of human communication to predict topology (green solid arrow) allows us to determine those features that best reflect this latent variable (green dashed arrow).}
    \label{fig:tie_strength}
\end{figure}

In this study, we assume that \emph{the strength of a tie is a latent variable expressed both in patterns of dyadic interactions and in network topology}, the latter following Granovetter's overlap hypothesis \cite{Granovetter}. This way, we use neighbourhood overlap as a benchmark that allows us to compare different characteristic features of communication events taking place on a tie: we use a feature's predictive capacity for neighbourhood overlap as proof of association with the latent strength of a tie. This approach is shown schematically in Figure \ref{fig:tie_strength}. 

This paper is structured as follows: first, we discuss how tie strength has been conceptualized and measured in previous research, both from the sociological and network-scientific perspectives. Then, we explore different modelling approaches to human communication which serve as a theoretical basis for our predictive features. We address the temporality of our data (a) as time series or sequences of interactions and (b) as events that occur within natural daily rhythms and weekly social cycles. Following this, 
we present results obtained with interpretable machine learning models and statistics for linking temporal features with neighbourhood overlap. We determine the importance of different variables as proxies for topological tie strengths, and show that combining multiple temporal features leads to most accurate predictions. We conclude with discussion.  

\section*{Measuring Tie Strength}
\subsection*{Historical background}
Despite its relatively intuitive definition in terms of emotional closeness \cite{Marsden1984}, the strength of a tie is a sociological concept with no direct indicator and its measurement requires prior theoretical definition and empirical validation \cite{Carmines1982, Marsden1984, Marsden2012, Wuchty2009}.
We can broadly distinguish two methodological variants: early studies often borrow from social psychology and rely on self-reported surveys, whereas more recent studies build on the availability of large sets of auto-recorded and behavioural data, which have spawned a wide array of methods and lines of research. 

The first conceptualizations of tie strength focused on intrinsic tie-level characteristics, such as relationship 'closeness' or kinship (e.g., relatives have strong ties while neighbors have weak) \cite{Marsden1984}. Alternatively, some researchers highlighted the effect of ties on the nodes, such as the provision of emotional support \cite{Wellman1982}, or the ability to handle multiple contexts \cite{Granovetter, Granovetter1985}. In his paper, Granovetter did not delve deeply into the definition of tie strength, characterizing it as a "possibly linear" combination of four constituting dimensions---time, emotional intensity, intimacy, and reciprocity \cite{Granovetter}. Many early studies analyzed social ties via standardized surveys that enquired about friendship, emotional support, frequency of contacts, or advise seeking \cite{Marsden1984, Wuchty2009, Carmines1982, Friedkin1990}, while acknowledging the limitations of  self-reported and, to a large degree, unilateral data for dyadic interactions \cite{Marsden1984, Friedkin1990, Brewer2000}. Marsden and Campbell \cite{Marsden1984} used survey data to determine which proxies for Granovetter's dimensions were most strongly associated with self-identified tie strength, suggesting that tie strength could be a multidimensional concept. 

Other lines of research have highlighted the temporal and dynamic aspect of human relationships, characterizing qualitative differences by relationship stages: initiation, maintenance and decay \cite{Wilmot1987, Dindia1993, Friedkin1990}. Gradual stages of reciprocity were identified as a key component in friendship formation \cite{Hallinan1978, Friedkin1990}. 
Burt \cite{Burt2000, Burt2002}, on the other hand, argued that factors associated with strong ties (homophily, social status, embedding, and inertia) are also associated with slower tie decay, but that tie decay is guided by nodes via selection processes and learning of social routines. Notably, both relationship initiation and decay were conceptualized as involving topological changes in social networks \cite{Wilmot1987}. Burt \cite{Burt2000} found evidence that embedded ties were associated with slow decay, but that disruptions in embedding implied faster decay. Some of these topological changes around tie decay were later be examined in dynamical contexts \cite{TimeAllocation, Backstrom2014, Navarro2017}. Moreover, even ties that were neither nascent nor decaying were established to be highly dynamic \cite{Dindia1993}, with Wilmot arguing that relational stability does not imply that relationships are static, but that there is a minimal agreement on about the relationship which is reflected on communication patterns \cite{Wilmot1987, Ayres1983}. 

From a socio-psychological perspective, Feld \cite{Feld1981} focused on how ties appear in social contexts that facilitate interaction, named \textit{foci}. 
Tie strength was thus theorized to be determined by sociological roles: a small and constraining focus (such as a nuclear family) might imply higher strength, but the interaction of multiple foci explains the multiplexity of ties, thus conveying the idea that two people interact in different contexts and social groups. The concept of foci was more recently exploited by \cite{Backstrom2014} to identify romantic partners, finding that people in romantic relationships have a focalized network structure---they both share a large number of common friends, but these friends belong to different foci, so they are \textit{dispersed}, or not connected among themselves. 

In the recent decades, technological advances and the advent of telecommunication devices and social media platforms have provided access to an unprecedented collection of auto-recorded data \cite{Kahanda2009}. This has generated new methodologies and conceptualizations of tie strength, which depend not only the underlying social network, but also on the data source with the appearance of distinct communication channels, such as phone calls or emails, but also of specific social platforms, such as Facebook, MySpace or Twitter. This has opened various previously unattainable lines of research, such as research on large-scale network properties \cite{Onnela2007}, characteristics of human communication networks \cite{Miritello2013b}, temporal networks \cite{JariTemp}, link prediction \cite{Wang2011}, and link decay \cite{Raeder2011}. In these cases, many studies adopted quantifications of tie strength in terms of communication intensity, such as the total number or duration of contacts \cite{Onnela2007, JariTemp, Wuchty2009, MiritelloStrategy, Navarro2017}. Other approaches have complemented the use of auto-recorded data with either surveys on emotional closeness \cite{Wuchty2011, Wuchty2009, Gilbert2009, Saramaki2014} or tagged human interactions in online platforms, such as interactions with  spouse or close friend \cite{Backstrom2014, Kahanda2009}, while other studies have determined features inspired by Granovetter's four dimensions; Navarro \cite{Navarro2017}, for instance, determined that strong ties were those that were unlikely to decay and identified features that predicted this. 

\subsection*{Tie strength and network topology} 

In this paper, we focus on untagged human interactions, where our goal is to infer the latent tie strength from behavioural features of communication. We conceptualize \textit{tie strength} as a latent variable that manifests independently in both network structure and communication patterns, so that strong ties are embedded in dense network communities while weak ties serve as inter-community bridges \cite{Granovetter, Granovetter1985}. 
The network structure and communication patterns are considered independent in the sense that no data used for computing network structure around the tie is used again for computing the temporal features of the tie's communication patterns.
Under our framework, the embeddedness or friendship overlap of a tie serves as a baseline that relates tie strength to features of communication. In this sense, variables with good predictive performance of topological features serve as better proxies for tie strength, at least to the extent they are reflected in local network topology. 

We measure embeddedness using 
topological overlap \cite{Onnela2007}, $O_{ij}$, which is defined as the Jaccard similarity of the sets of neighbors of two nodes $i$ and $j$, a measure that can be interpreted as the percentage of common neighbors around a tie. Formally the topological overlap is defined as
\begin{equation}
O_{ij} = \frac{|\mathcal{N}_i \cap \mathcal{N}_j|}{|\mathcal{N}_i \cup \mathcal{N}_j|}\,,
\end{equation}
where $\mathcal{N}_i$ is the set of neighbors of node $i$. The Granovetter effect---the increase in embeddedness along with tie strength---was previously observed using overlap and number of calls ($w_{ij}$) and total call time ($l$) as proxies for link weights \cite{Onnela2007, MiritelloStrategy}. Previous research has also found that different communication patterns entail topological changes in social networks \cite{Wilmot1987, Navarro2017, MiritelloStrategy, Backstrom2014}, and indeed tie evolution has also been associated with distinct features of human communication for both tie creation and decay \cite{Navarro2017}. 

On this matter, our focus is not on detecting topological change. This is because (i) topological variations have been shown to occur over long periods of time \cite{MiritelloStrategy, Navarro2017, Backstrom2014} which correspondingly requires long-term longitudinal data and (ii) they entail the additional problem of uncoupling bursty communication patterns from changes in the underlying social structure \cite{Miritello2013b, Navarro2017}. This issue is heightened by different social strategies empirically observed in communication networks, where \textit{explorers} display a large turnover of weaker ties, while \textit{keepers} prefer a smaller circle of stronger ties \cite{MiritelloStrategy}. In addition, we know overlap to be a decreasing function of the aggregation window for communication networks \cite{Krings2012}. To address these problems, we assume that tie strength remains constant during our observation period, which we expect to be true for most relationships in a span of a few months \cite{Wilmot1987, Ayres1983, Dindia1993}, and provide a dynamic measure of overlap that penalizes ties that are not active over a long period. We measure overlap in a dynamic manner by establishing a smaller aggregation window, $\Delta T$, which we shift over the full period and to obtain a time series of overlap values $\{O_{ij}^t\}_{t=1}^{N_t}$. We use the average of our time series as a measure of temporally averaged overlap, $\Ot_{ij}$.
This variable emphasizes edges that are relatively close in time. 
We obtained dynamic overlap with $\Delta T = 1$ month, which we justify since empirical evidence on similar datasets \cite{Krings2012} has found overlap to become relatively stable at an aggregation window of this size. To serve as a baseline, we repeated the same experiments using the static overlap over the full observation on the SI. 

\subsection*{Source Data}
We use a Call Detail Records (CDRs) database from a single operator in a European country \cite{Onnela2007}, with an observation period of four months during 2007 and a market share of 20\%. CRDs are communication logs recorded by mobile service providers, where basic information of the interaction is sequentially stored, including, e.g., caller, callee, timestamp and duration. CDRs from single operators are functionally a statistical sample of a complete dataset of interactions \cite{Onnela2007}. Despite lacking a full network, our dataset does provide full ego networks centered on our operator's subscribers. We thus focus our study on the strength of ties that fully belong to our operator (both nodes in a tie are subscribers), involving $\sim 6.5$ million nodes and $\sim 26.4$ million ties; however, for network topology we also use ties of non-company users, which correspond to an additional $\sim 76$ million nodes and $\sim 530$ million ties. This methodological choice guarantees that there is no bias related only single operator links being included in the overlap calculation. This  mitigates the concerns that our dataset is not a random sample - because family ties, friendship recommendations and regional differences in market share may be drivers when customers choose a mobile service provider, and these differences might result in biased estimates of overlap. 

\begin{figure}
    \centering
    \includegraphics[width=.9\textwidth]{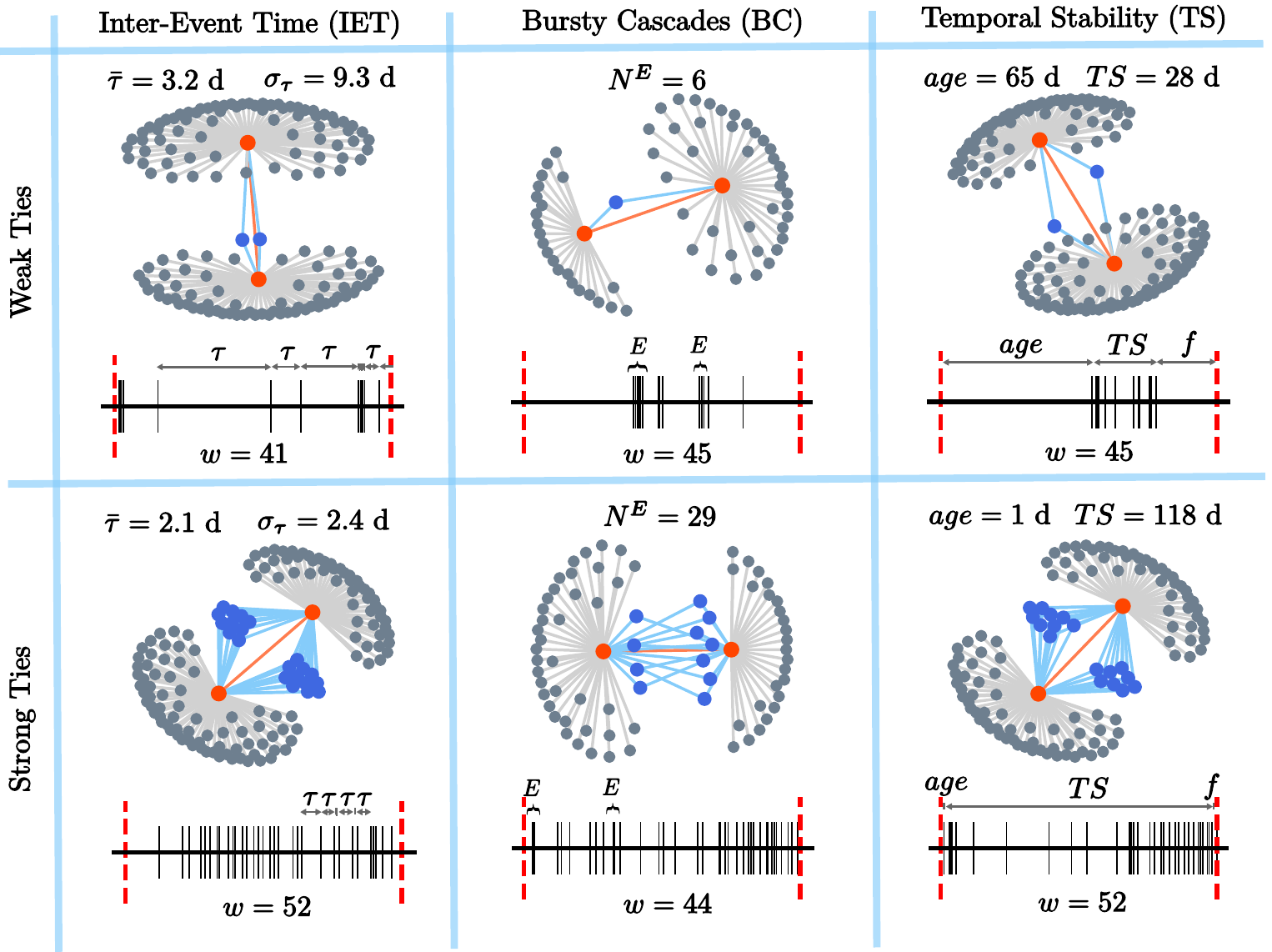}
    \caption{Differences in neighbourhood overlap (\textit{top} and \textit{bottom}) and selected temporal features of human communication (columns) for ties of similar communication intensity ($w \in [40, 55)$). Weak (\textit{top}) and strong (\textit{bottom}) ties are defined by low and high overlap, respectively, where we analyze communication patterns of red ties ($ij$), blue ties represent common neighbors ($\mathcal{N}_i \cap \mathcal{N}_j$), and grey ties represent neighbors of either $i$ or $j$ but not both. (\textit{Left}) IET distribution; we model the time between two consecutive calls as a homogeneous process via a random variable $\tau$, and obtain statistics on the IET distribution, such as $\bar{\tau}$ and $\sigma_{\tau}$, both measured in days. (\textit{Center}) Bursty Cascades; given a parameter $\Delta t$, we identify event \textit{bursts} $E$ as sub-sequences of calls that are placed at most $\Delta t$ seconds within each other. $N^E$ is the number of events. (\textit{Right}) Temporal stability; we focus on the first and last events and their distribution within the observation period (red dashed lines), and determine the $age$ as the time until the first call, the temporal stability $TS$ as the time window where we observed events, and freshness $f$ as the time between the last event and the end of the observation period.}
    \label{fig:sequential_modelling}
\end{figure}

\section*{Features of Human Communication}
\label{sec:tie_strengths}

Our aim is to determine features that might encode information on the tie strength not captured by intensity variables. Figure \ref{fig:sequential_modelling} illustrates this idea by showcasing ties in our data of similar communication intensity $w$ that differ both in overlap and communication patterns. 
In the following, we expand on these temporal features of human communication, and use them as predictors of overlap, comparing them with the widely-used number of contacts as a communication-intensity measure. 

A key assumption of this work is that differences in the strengths of ties are reflected in communication patterns of dyadic interactions. 
Based on these data, we collected variables from existing literature that model different aspects of human communication and developed some new indicators. We roughly divide our approach in two: measures building on the sequential nature of our data and measures  focusing on daily and weekly behaviour. 

\subsection*{Intensity Features of Human Communication}

Features related to communication intensity have commonly been used as a proxy for tie strength \cite{Onnela2007, Saramaki2015, Krings2012}. We denote the number of calls as $w$, as this is commonly used as link weight in social network analysis. We further analyze communication intensity in terms of total call time $l = \sum_{i=1}^w l_i$ where call $i$ has length $l_i$, as well as average call time $\hat{l} = l / w$. 

We also characterize the reciprocity $r$ as an intensity feature \cite{Navarro2017, Hallinan1978}, which we measure via 
\begin{equation}
r_{ij} = \left|\frac{\vec{w}_{ij}}{w_{ij}} - \frac{1}{2} \right|
\end{equation}
where $\vec{w}_{ij}$ is the number of calls placed by $i$ to $j$, so that $r \approx 0$ implies that both users placed a similar amount of calls, while $r\approx 0.5$ reflects an imbalance.

\subsection*{Sequential Features of Human Communication}

At the level of ties, CRDs record a time series of events. Most of the measures based on these time series are based on the intuition that regular contacts are more significant than for example brief periods of large contact intensity; we exemplify some of these modelling approaches in Figure \ref{fig:sequential_modelling}.
In this section, we first focus on measuring the number of time periods during which the tie has been active.
Second, we consider the time elapsed between consecutive calls via the inter-event time (IET) distribution. Third, we focus on correlated bursty behaviour and memory effects, using the distribution of event \textit{bursts}. Last, we focus on behavioural changes within the observation window, with variables that have been previously associated with tie creation and decay.

\subsubsection*{Counting active periods}
The regularity of a time series can be measured by counting the active periods, such as hours or days, with at least one contact. We record the number of hours and days with events, $a_h$ and $a_d$, respectively.  Since we know human communication to be bursty \cite{KarsaiBursty, Goh2008, Barabsi2005, KarsaiSmall}, this aggregating process serves to remove temporal correlations to different degrees. These variables also allow for the incorporation of different communication channels, such as phone calls and text messages \cite{Heydari2018}.

\subsubsection*{Inter-event time distribution}

We measure the IET, the elapsed time between consecutive calls ($\tau$), depicted in Figure \ref{fig:sequential_modelling} (\textit{left}). Given the set of interaction times $\{t_0, t_1, \ldots, t_n\}$, we obtain the $k$th inter-event time $\tau_k = t_{k+1} - t_k$, and in practice may estimate moments from this distribution from the empirical observations $\{\tau_k\}$. 

The IET distribution encodes uncorrelated information about the times between consecutive calls. This uncouples temporal correlations between events \cite{Jo2019Burst} while discarding possible memory effects between consecutive inter-call times. This allows us to obtain general call patterns such as the mean IET $\bar{\tau}$ and the standard deviation of the IET $\sigma$, where a small $\bar{\tau}$ would imply more frequent communication, which has been theorized to occur when ties are strong \cite{Wilmot1987}. Previous research has estimated the IET distribution to be heavy tailed 
\cite{Goh2008, Kivela2015} and bursty, so that short spikes of activity are followed by long periods of inactivity \cite{Goh2008, KarsaiBursty}. In this sense, the IET distribution provides a natural way to characterize uncorrelated burstiness via the burstiness coefficient $B = \frac{\sigma - \bar{\tau}}{\sigma + \bar{\tau}}$ \cite{Goh2008}, which takes value $B=-1$ for completely regular IETs, $B=0$ for Poissonian behaviour, and $B=1$ for completely bursty or irregular behaviour. A related measure is the average relay or waiting time $\tau_R$, which is defined as the time between a random point in time and the next event. It can be used as a local measure of the speed of information spreading over the link, and when normalised with $\bar{\tau}$ it has been shown to be a non-linear function of $B$ \cite{Kivela2012}. 

\subsubsection*{Bursty Cascades}

Temporal correlations, neglected by the IET distribution, are common in human communication \cite{KarsaiBursty}. Our next set of features places a larger importance on bursts as determined via a parameter $\Delta t$. Karsai \emph{et al.} \cite{KarsaiBursty} define a bursty cascade by the number of consecutive communication events $E$ that took place  within a time period of $\Delta t$ or less; in other words, events $k$ and $k+1$ are part of the same bursty train iff $\tau_k = t_{k+1} - t_k \leq \Delta t$, as depicted in Figure \ref{fig:sequential_modelling} (\textit{b}). 

This approach been used to find that $P(E)$, the distribution of the number of events in a bursty cascade, is also heavy-tailed over a range of $\Delta t$ values \cite{KarsaiBursty,Jo2019Burst}. In contrast, if the event times are uncorrelated but follow the same IET distribution, there is an exponential decay for $P(E)$.
The structure of correlations that can be constructed from the bursty cascades at different resolutions $\Delta t$ is completely independent of the IET distribution \cite{Jo2019Burst}.
This allows for a flexible characterization of human communication, where the main focus is not on calls, but on call cascades. In this respect, this shift of focus provides new features of communication frequency via the number of cascades, but also via how calls are distributed within cascades. 

We use a set of variables related to bursty cascades, including the mean number of events per cascade $\bar{E}$, the standard deviation $\sigma_E$, the coefficient of variation $CV^E = \frac{\sigma^E}{\bar{E}}$ and the number of bursty cascades $N^E$. We chose to use $\Delta t = 26$ hours, since preliminary tests showed that this yielded the best association with overlap. These results, available in the SI, corroborate that $P(E)$ is not overtly sensitive to the choice of $\Delta t$. 

\subsubsection*{Temporal Stability}

The above approaches implicitly assume that behaviour doesn't change in time, that is, they measure communication activity while assuming that the underlying social relationship remains constant. As previously stated, it is not trivial to disentangle bursty communication patterns from the underlying dynamic relationship, where long IETs might be interpreted as tie decay \cite{Miritello2013b}. We may, however, measure behavioural changes during the observation window, for which we use two sets of variables. For the first set of variables, we divide the observation window into three sub-intervals, measuring a) the $age$ of a tie as the first observed communication event \cite{Holme2015} b) the temporal stability ($TS$) of a tie as the elapsed time between the first and last communication events, and c) the freshness of a tie $f$ as the time elapsed between the last communication event and the end of the observation window
\cite{Gilbert2009, Raeder2011, Holme2015, Navarro2017}. For the last variable, we use relative freshness $f^r = f/\bar{\tau}$, which allows us to compare the time elapsed with no communication with the average IET, a metric which has been used to predict tie decay \cite{Navarro2017}. 

\subsubsection*{Distribution of Bursty Cascades}

Next, our goal is to characterize \emph{when} communication takes place within the observation window, in a similar fashion to temporal stability features. The previous measures, however, used only the first and last communication events, while we will now work on the whole set of interactions. We decouple correlated bursty behaviour by focusing on the distribution of bursty cascades within the observation period, as opposed to the distribution of calls.

We define our variables as follows: given a parameter $\Delta t$ and a sequence of interaction timestamps $\{t_j\}_{j=1}^{w}$, where each $t_j$ has been normalized to the interval $[0, 1]$  defined by the observation window, we obtain a sequence of timestamps for bursty trains $\{t^*_i\}_{i=1}^{N^E}$, where $t^*_i$ corresponds to the first observed event within bursty train $i$. We define the average interaction time $\bar{t} = \frac{1}{N^E}\sum_i t^*_i$, and the associated standard deviation $\sigma_t$. We found that overlap decreases for average interaction times that were skewed on the observation window (average values $\bar{t}_{ij}$ far from $t=0.5$). For this reason, we included a feature that measures deviation from $t=0.5$ as a test statistic for difference of means with unknown variance $T = \frac{\bar{t} - 0.5}{\sigma_t \sqrt{N^E}}$. We use $\log(T)$ to penalize outliers. 

\subsection*{Daily and Weekly Features}

Human behaviour is regulated by the interplay of natural and social factors that determine different degrees of activity during, e.g., the day-night cycle or weekday-weekend cycle \cite{Panda2002, Aledavood2017, Wittmann2006Social,aledavood2015a}. Our goal in this section is to determine whether these fluctuations are also reflected in network topology. We focus on two main sets of variables: first, we analyze differences in daily activity patterns, and second, differences in call profiles during the week. 

\subsubsection*{Differences in Daily Patterns}

Although humans typically follow 24-hour cycles determined by daylight, behaviour during these cycles has been found to be highly heterogeneous \cite{Aledavood2015b,Aledavood2015c}. In particular, there are prominent individual differences among the morningness or eveningness of people \cite{Aledavood2017, Adan2012}; that is, the propensity to be more active during the morning or evening. We look for differences in daily call patterns of people forming dyads, and use these as a candidate measure for predicting tie strength. This variable is conceptually different from the previous ones as it is defined using information from two nodes instead of a single tie. Our hypothesis is that there are several reasons why people linked by strong ties have more similar daily call rhythms: people might have habitual calling patterns, the activities of friends might be synchronized through joint activities, or there might be latent drivers of call behaviour that are also associated with homophily, such as age.

For each person, we compute a 24-hour daily distribution $P = (p^0, \ldots, p^{23})$ of the fraction of outgoing calls placed during each hour. For each tie, we then measure differences in the daily distributions by using the Jensen-Shannon Divergence (JSD), chosen for its ability to handle zero-valued probabilities. The JSD is defined for two discrete probability distributions $P_0$ and $P_1$ as
\begin{equation}
JSD(P_0, P_1) = H\left(\frac{1}{2}P_0 + \frac{1}{2}P_1 \right) - \frac{1}{2}\left(H(P_0) +H(P_1)\right),  
\end{equation}
where $H$ is the Shannon entropy, $H(P) = - \sum_t p(t)\log\left(p(t)\right)$. 

\subsubsection*{Weekly Activity Profiles}

Our last focus is weekly behaviour, where we identify times during the weekly cycle where a distinct call profile might be associated to higher/lower topological overlap. The motivation is that ties within different groups or foci might be associated with different call-placing patters: activity between colleagues can be expected to differ from that between family members or friends \cite{Feld1981}. We follow a two-step procedure where we first divide the week into $7\times 24=168$ hourly bins, and to each bin we assign the fraction of calls placed by both nodes in a tie. Unlike for the daily patterns, the focus is therefore on ties instead of node-level behaviour. This high-granularity approach yields features that are too sparse to be interpretable; for this reason, as a second step we perform dimensionality reduction based on the overall call profiles of the whole dataset. We base this dimensionality reduction on our 168-feature correlation matrix and their association with overlap. For details, see SI. 

\section*{Results}
\label{sec:results}
\subsection*{Clustering of weekly call patterns}

\begin{figure}
    \centering
    \includegraphics[width=.65\textwidth]{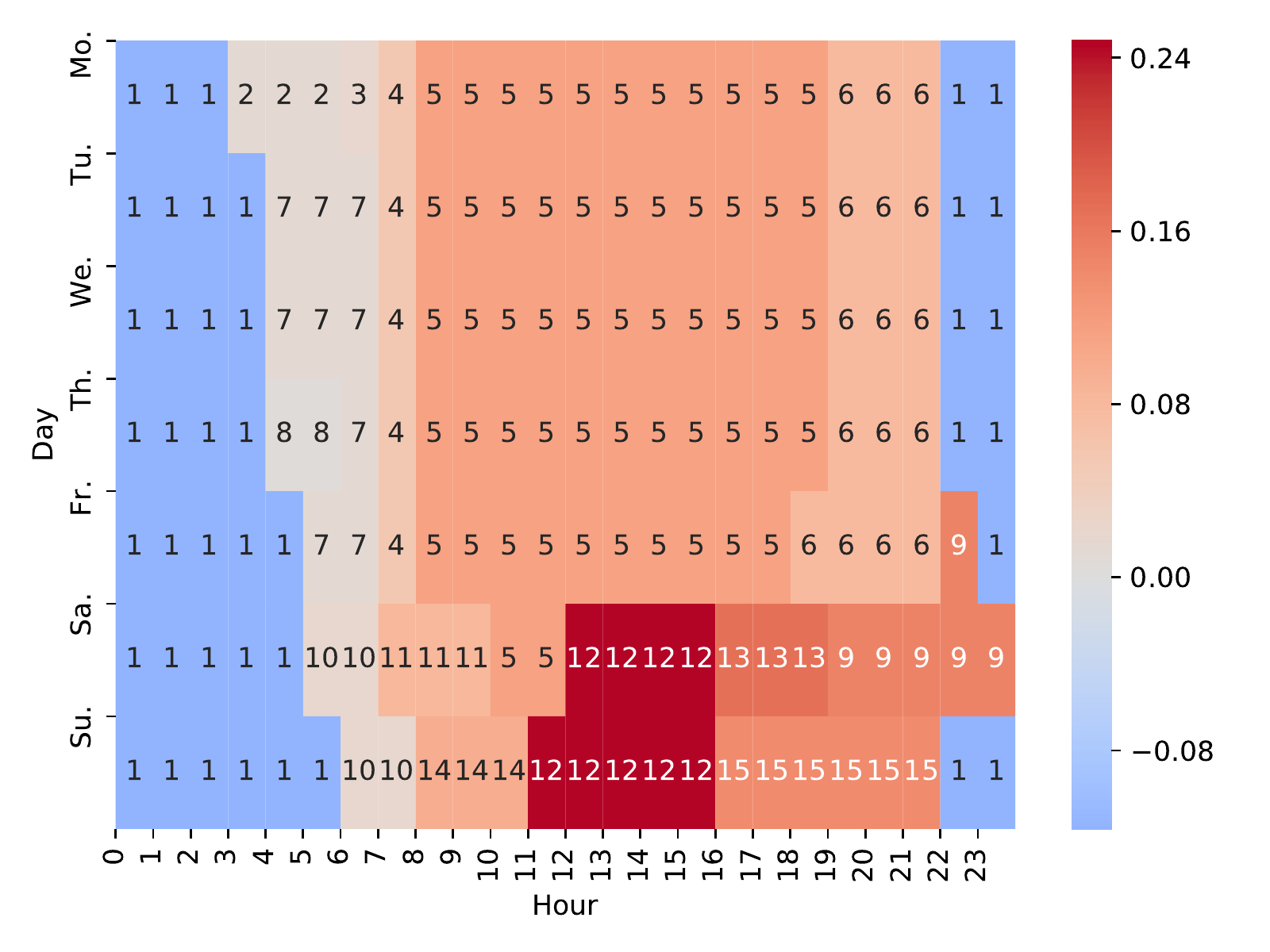}
    \caption{Composition of weekly call profiles  of social ties and their association with the neighbourhood overlap. Each bin represents an hour of the week (x-axis: hours, y-axis: days), and the number inside the bin is the corresponding cluster index. The bin's color indicates the level of Pearson's correlation of each tie's fraction of events in the bin's cluster with the overlap. \emph{E.g.}, a tie's topological overlap correlates positively with the fraction of calls across that tie that take place between noon and 4 PM on weekends.}
    \label{fig:clusters}
\end{figure}

Figure \ref{fig:clusters} depicts our results on how different weekly call profiles are associated with different overlap values. After our dimensionality reduction process, we obtained 15 clusters $\{Ci\}_{i=1}^{15}$ which constitute a weekly call profile vector $C^\ast$ for each tie; we normalize the component contributions so that  $\left|C^\ast_{ij}\right|=1$. We find that there is heterogeneity in the association between the call profiles and overlap: the fraction of a tie's calls that belong to cluster $C_{12}$ (weekend late morning and early afternoon) correlates positively with the overlap, whereas there is a low negative correlation for late-night calls (cluster $C_1$). 


\subsection*{Predicting overlap from tie features}

\begin{figure}
    \centering
    \includegraphics[width=.95\textwidth]{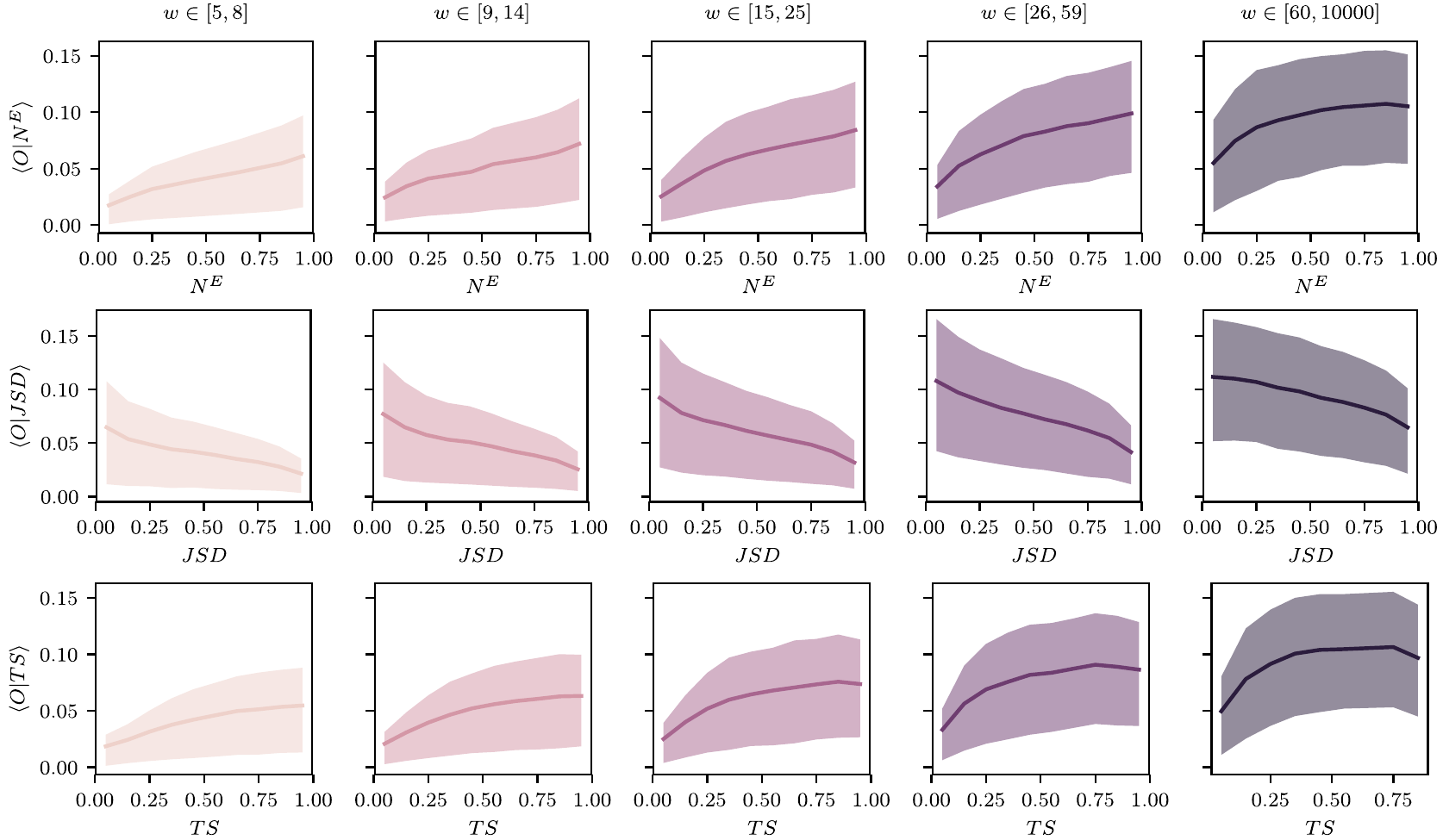}
    \caption{Average topological overlap given the ranks of three variables correcting for different levels of communication intensity ($w$), with the shaded area depicting 80\% of the distribution. From top to bottom: number of bursty trains ($N^E$), Jensen-Shannon Divergence for difference in daily patterns ($JSD$), temporal stability ($TS$). Variable rankings are normalized to be on the $[0,1]$ interval.}
    \label{fig:fivevars}
\end{figure}

Our goal is to predict topological overlap using  features computed for ties, and to compare their performance to simple communication intensity measures.
Table \ref{tab:1} contains a list of the features used in our study. First, to show that such features have explanatory power beyond communication intensity $w$, we have stratified ties into groups based on $w$ and studied how the overlap depends on the variable associated with each feature within the groups. This dependence is shown for three features---the number of bursty trains ($N
^E$), the daily pattern difference ($JSD$), and the temporal stability $TS$---in Fig.~\ref{fig:fivevars}. It is clear that these features correlate with overlap even within groups of ties with a narrow intensity range; this holds for other measures of communication intensity and other features (IET, etc) as well. See SI for further details.

\begin{table}[h!]
\caption{Features of human communication used in our analysis. Our feature types - Intensity (I), Active Periods (AP), Inter-event time (IET), Temporal Stability (TS), Bursty Cascades (BC), Distribution of bursty cascades (DBC), differences in daily patterns (DP) and clusters for weekly activity.}
      \begin{tabular}{ccc||cc}
        \hline
        Type & Variable  & Name  & Cluster & Description\\ \hline
        I &  $w$       & Number of calls & $C1$ & Late night and early morning\\
        I &  $l$       & Total call duration & $C2$ & Monday early morning\\
        I &  $\bar{l}$ & Average call duration & $C3$ & Monday early morning\\
        I & $r$ & Reciprocity & $C4$ & Weekday 7 am\\
        AP & $a_d$     & Number of days with calls & $C5$ & Weekday afternoon \\
        AP & $a_h$     & Number of hours with calls & $C6$ & Weekday evening\\
        IET &  $\bar{\tau}$ & Mean IET & $C7$ & Weekday earkly morning\\
        IET &  $\sigma_{\tau}$ & Std. Dev. of IET & $C8$ & Thursday early morning\\
        IET &  $B$ & Burstiness Coefficient & $C9$ & Weekend evening \\
        IET &  $\bar{\tau}_R$ & Average Relay Time  & $C10$ & Weekend morning \\
        TS  &  $\hat{f}$ & Relative freshness & $C11$ & Saturday Morning \\
        TS &  $age$ & Age & $C12$ & Weekend afternoon\\
        TS &  $TS$ & Temporal Stability  & $C13$ & Saturday late afternoon \\
        BC &  $N^E$ & Number of busty events  & $C14$ & Sunday morning \\
        BC &  $\bar{E}$ & Average calls per bursty event & $C15$ & Sunday afternoon\\
        BC &  $\sigma^E$ & Std. Dev. of event distribution & $C^*$ & Vector of clusters\\
        BC &  $CV^E$ & CV of event distribution  & & \\
        DBC &  $\bar{t}$ & Avg. interaction time & & \\
        DBC &  $\sigma_t$ & Std. Dev. of interaction times  & &\\
        DBC &  $\log(T)$ & Test statistic for avg. interaction time  & &\\
        DP & $JSD$  & Differences in daily behaviour  & & \\ \hline
      \end{tabular}
      \label{tab:1}
\end{table}

For our predictive task, we applied machine-learning models (see below) to three different scenarios: a) using each feature as a single predictor, b) using each feature along with the best-performing features in the previous task, and c) using the full set of features. These scenarios allow us i) to identify the individual features that encode most information on overlap, ii) 
to compare the performance of these features with commonly used measures and see how complementary they are,
and iii) to assess the maximum predictive capacity of our features and to know their relative importance. 


As there is no natural scale for overlap that would relate it to the latent tie strengths, we take a nonparametric approach and focus on predicting overlap rank instead of overlap values. The prediction problem itself was transformed into the binary decision problem of predicting high/low overlap values.  We selected a range of high/low overlap values $\{\Ot_{\alpha}\}$ according to the overall distribution, with cutoff points every fifth percentile $\alpha$. For each scenario, we ran four machine-learning models that allow for interpretable results: logistic regression (LR), random forests (RF), quadratic discriminant analysis (QDA) and AdaBoost classifier (ABC). We obtained a sample of 500,000 ties, performed 3-fold cross-validation for our overlap prediction tasks, and measured the predictive performance of our models via Matthews Correlation Coefficient (MCC) \cite{Matthews1975}, a classification performance metric for binary data related to Person's correlation coefficient, and used for it's ability to handle imbalanced and asymmetric data \cite{Boughorbel2017}. 

\begin{figure}
    \centering
    \includegraphics[width=.9\textwidth]{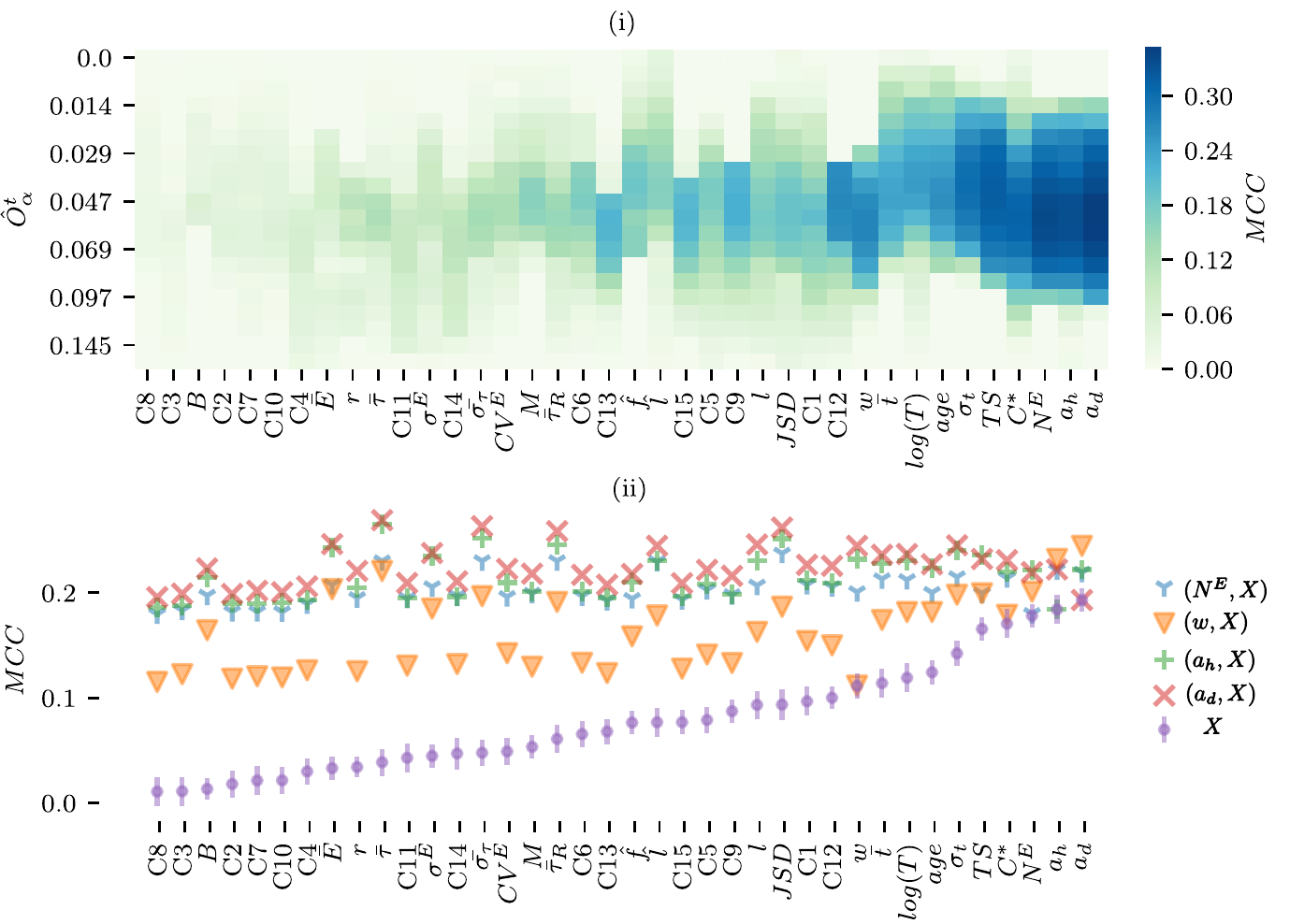}
    \caption{Matthew's Correlation Coefficient (MCC) for dynamic overlap prediction for different scenarios. The $x$-axis represents features used for prediction in the machine learning models. (\textit{i}) Maximum MCC for four models trained with single-feature predictors, where each variable is used to predict static overlap using RF, ABC, LG and QDA. The $y$-axis represents the averaged overlap $\hat{O}^t_{\alpha}$ cutoff value for binary high/low overlap classification where $\alpha$ increases every 5th percentile. The color represents the maximum $MCC$ over the four ML models. Variables are ranked according to their average performance over all cutoff values $\alpha$. (\textit{ii}) Comparison between single and dual-variable models $(F, X)$ for $F = N^E, w, a_h, w_d$, where we depict the average performance over all $\alpha$s. For the single-variable case ($X$), error bars are 2 standard deviations, obtained via bootstrapping.}
    \label{fig:dynamic}
\end{figure}

The predictive performance of all individual features is shown in Figure \ref{fig:dynamic}. 
Results are shown for the averaged overlap, $\Ot$. For $O$, see SI. In addition, for the single and dual feature scenarios, we include $C^* = (C1, \ldots, C15)$, the vector of cluster weights for a tie's weekly call profile. Although $C^*$ is not a single variable, we include it as a means of comparing how much information is encoded by the weekly call profile, and include a full analysis of $C^*$ in the SI. 

On average, nine features outperform the number of calls $w$ in predicting topological overlap:
the number of days $a_d$ and the number of hours $a_h$ with calls, the number of bursty trains $N^E$, temporal stability $TS$, the weekly call profile $C^*$, three features of the distribution of bursty cascades (DBC), and tie $age$. 

The performance of predictors differs for low or high overlap cutoff values $\Ot_{\alpha}$, which is indicative of how these measures perform overall: $a_d$, $a_h$ and $N^E$ encompass a broad spectrum of values centered around the median of the overlap distribution. The weekly call profile $C^*$ has a wider spectrum and is one of the few features with nonzero MCC for all $\alpha$ values, even though its predictive performance for mid-range values of $\alpha$s is smaller than that of the three top-ranking features. The features TS and DBC ($TS$, $age$, $\sigma_t$, $\log(T)$ and $\bar{t}$) tend to have higher predictive performance skewed towards smaller $\Ot_{\alpha}$ values. 

The fraction of calls in some component clusters of the weekly call profile is surprisingly predictive of overlap. In particular the weekend day cluster alone ($C12$) has a high predictive performance for mid-range values of $\Ot_{\alpha}$. The cluster for early morning and weekday nights ($C1$) also ranks highly for average overlap prediction. In this case, correlation with overlap was mostly negative, suggesting that a high fraction of calls at certain times might indicate weak ties. We provide a more complete analysis of the predictive power and the  importance of the different components of $C^*$ in the SI.

We compare the effect of including additional information on the prediction task by using pairs of variables as predictors $(F, X)$, where $F$ is one of the three best-performing features ($a_d$, $a_h$, $N^E$) or the number of calls ($w$), and $X$ is the set of all other features. These variables' performance increases moderately when used in tandem when compared with the baseline single predictor, with an average increase of 16.8\% for $(a_d, X)$ against $a_d$; for a small set of features, however, the average performance increases considerably, up to 39.5\%  for $(a_d, \bar{\sigma})$. Notably, the compound effect of feature pairs is higher with variables that have low single-feature predictive performance. This includes variables derived from the IET, such as $\bar{\tau}$, $\bar{\sigma}_{\tau}$, differences in daily patterns $JSD$, and features of call duration, $l$ and $\hat{l}$. 

\begin{figure}
    \centering
    \includegraphics[width=.9\textwidth]{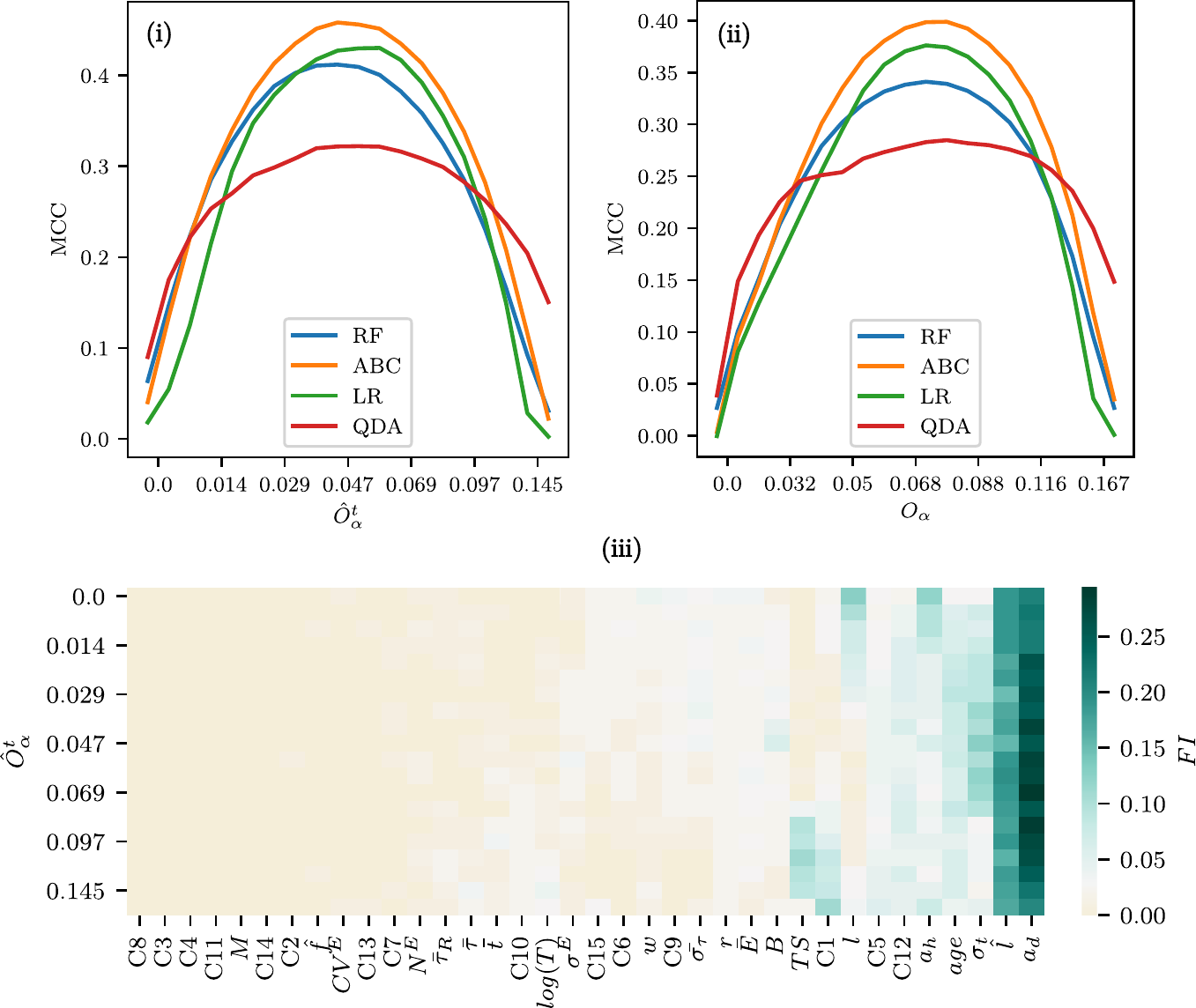}
    \caption{Full model scenario. (\textit{top}) MCC for four different models used in prediction of (\textit{i}) dynamic and (\textit{ii}) static overlap. (\textit{iii}) Feature importance (FI) for the overall best performing model, ABC, for prediction of dynamic overlap. Features are ranked by their average importance over all cutoff values $\alpha$.}
    \label{fig:full_scores}
\end{figure}

Last, we used the full set of features in the overlap prediction task, with the aim of obtaining  maximal predictive performance and understating the relative feature importance (FI), defined as the mean decrease in impurity induced by a feature \cite{Louppe2013}. Figure \ref{fig:full_scores} displays the maximum MCC for static and dynamic overlap using different models. Both cases follow a similar trend where the best predictive performance is achieved roughly in the middle of the distribution, and where the ML models RF, ABC and LR achieve similar results. Notably, all models perform slightly better for the averaged overlap $\Ot$, with maximum MCC of $0.457$, as opposed to static overlap $O$, with maximum MCC $0.399$. 
This can be an indication of the averaged overlap $\Ot$ being a better proxy for the latent tie strength as discussed earlier.
The performance of model QDA is noticeably different from the other three models, outperforming all models for extreme overlap cutoff values but displaying a notably flatter performance curve. As for feature importance, the effect of $a_d$ and $\hat{l}$ dominates other variables in our full model, which is characterized as having skewed FI values. This suggests that a high-performing model can be achieved with a subset of variables, or that some of these variables might contain redundant information on network topology. Many of the variables with high FI, however, correspond to widely different modelling approaches (AP, I, DBC, TS, $Ci$), suggesting that the interaction between network topology and behavioural features is multifaceted. 

\section*{Discussion}

Human communication patterns encode information on their local network topology. In this paper, we conceptualized tie strength as a latent variable that manifests independently as both local network topology and as patterns of communication between two nodes. We identified which features of these dyadic interactions are the best predictors for the neighbourhood overlap of a tie and therefore for the latent tie strength. We find that while commonly-used aggregated measures such as the total number of calls are adequate indicants of network overlap, our results show that alternative proxy measures contain information not captured by mere intensity features. We focused on quantifying different temporal aspects of human communication, using both sequential and cyclical features. We showed that several of these distinct approaches capture information on network topology. 

The number of days and hours with contacts ($a_h$ and $a_h$, respectively) outperformed all other variables in the prediction task, as did the number of bursty cascades $N^E$. Notably, these variables are conceptually similar to features measuring communication intensity, but with the key difference that part of bursty behaviour is removed through temporal aggregation.
In addition to these, simple variables related to the time of the first and last communication ($TS$ and $age$) performed better than the communication-intensity features.

We introduced a weekly call profile $C^*$ and found it to be highly informative on the neighbourhood overlap of ties. Notably, even though $C^*$ was not the best predictor, it had the highest predictive power for the widest range of overlap cutoff values, providing a richer characterization that other features. Interestingly, $C^*$'s performance does not increase significantly in combination with new variables, which might suggest that the weekly profile contains information on intensity as well. A simple mechanism for encoding a large number of communication events could be through several active clusters, for example. What is more, we found strong evidence that individual calling times during the week convey information on network topology. Notably, for our dataset in a European country, weekend afternoons proved to have a higher correlation with overlap than most other variables, whereas weekday nights and early mornings were associated with low overlap. These results pave the way for interesting lines of research. For example, one can use different data sets to compare the differences of weekdays and times of days across contexts and cultures.

In the case of modelling bursty trains, the parameter $\Delta t$ determines the period where two calls are considered to be correlated. Previous research had found that the distribution of calls within trains did not vary significantly with different $\Delta t$ values \cite{KarsaiBursty}. Although we did find differences in predictive performance, which included an optimal value of $\Delta t^* = 26$ hours, we also found evidence that a wide range of $\Delta t$ values outperformed $w$. This suggests that in practical applications, the aggregation of temporally correlated calls might already improve the topological information encoded in the variable.

Measuring differences in daily call patterns ($JSD$) also proved to encompass topological information, an effect more evident when predicting static overlap (see SI) and dynamic overlap in our dual-variable scenario. This was slightly surprising, as the relationship to network topology is not as straightforward as other features. We hypothesized two possible explanations for this, which are not mutually exclusive. In the first case, there could be a latent homophilic effect, where activity encodes information on, \emph{e.g.}, age or work relations. A second possible explanation is that strong ties engage in correlated call events, where person A's call is followed by the person B's call. Despite the strong association, further research is needed to uncover the drivers of this relationship. The use of temporal stability also provides a useful characterization, as it is one of the most simple features that only requires two observations. Indeed, we do not delve into the effect of the observation window into the use of this variable, where tie decay is more likely to occur, along with the topological changes it implies \cite{MiritelloStrategy, Navarro2017}.

If one needs to pick a single simple measure for tie strength based on this study it would be the number of days with contact. However, this measure would have only about two thirds of the predictive power  as compared to using the full contact sequence (when measured with MCC to predict $\Ot$). That is, the latent tie strength is a combination of multiple features which reflects the different facets of human relationships. Our results suggest that such important facets include regularity of contact, total amount of time spent, and the type of interaction reflected by the time and weekday of the contact. 

We should also note here that we did not investigate the direction of causality, but only the association of variables. That is, we do not answer the question of if high overlap values are followed by high latent tie strengths or the other way around.  If each feature represents different aspects of the latent tie strength then one could also study each of them separately as predictors of overlap in the future \cite{Navarro2017} or vice versa. Moreover, our results might be dependent on cultural features, communication medium, technology and other variables, and thus might not be directly transferable to other data sets. However, if one has access to a social network based on contact events, then it is straightforward to use the framework we have set up here and find the features which are most important in a specific context.

Lastly, the list of features we constructed here is by no means exhaustive and it is based on the current literature on analysing temporal social networks. However, our framework provides a way to benchmark any new features as an independent predictor of the latent tie strength, or as an additional facet of the tie strength by inspecting its performance together with other features.

\bibliographystyle{unsrt} 
\bibliography{references}      

\begin{thebibliography}{10}

\bibitem{Onnela2007}
J.-P. Onnela, J.~Saram\"{a}ki, J.~Hyv\"{o}nen, G.~Szab{\'{o}}, D.~Lazer,
  K.~Kaski, J.~Kert{\'{e}}sz, and A.-L. Barab{\'{a}}si.
\newblock Structure and tie strengths in mobile communication networks.
\newblock {\em Proceedings of the National Academy of Sciences},
  104(18):7332--7336, April 2007.

\bibitem{Candia2008}
Juli{\'{a}}n Candia, Marta~C Gonz{\'{a}}lez, Pu~Wang, Timothy Schoenharl, Greg
  Madey, and Albert-L{\'{a}}szl{\'{o}} Barab{\'{a}}si.
\newblock Uncovering individual and collective human dynamics from mobile phone
  records.
\newblock {\em Journal of Physics A: Mathematical and Theoretical},
  41(22):224015, 2008.

\bibitem{Saramaki2015}
Jari Saramäki and Esteban Moro.
\newblock From seconds to months: an overview of multi-scale dynamics of mobile
  telephone calls.
\newblock {\em The European Physical Journal B}, 88(6), 2015.

\bibitem{Borgatti2009}
S.~P. Borgatti, A.~Mehra, D.~J. Brass, and G.~Labianca.
\newblock Network analysis in the social sciences.
\newblock {\em Science}, 323(5916):892--895, feb 2009.

\bibitem{KarsaiSmall}
M.~Karsai, M.~Kivel\"{a}, R.~K. Pan, K.~Kaski, J.~Kert{\'{e}}sz, A.-L.
  Barab{\'{a}}si, and J.~Saram\"{a}ki.
\newblock Small but slow world: How network topology and burstiness slow down
  spreading.
\newblock {\em Physical Review E}, 83(2), 2011.

\bibitem{Gonzalez2008}
Marta~C. Gonz{\'{a}}lez, C{\'{e}}sar~A. Hidalgo, and Albert-L{\'{a}}szl{\'{o}}
  Barab{\'{a}}si.
\newblock Understanding individual human mobility patterns.
\newblock {\em Nature}, 453(7196):779--782, jun 2008.

\bibitem{Li2019}
Tracey Li, Jesper Dejby, Maximilian Albert, Linus Bengtsson, and Veronique
  Lefebvre.
\newblock Estimating the resilience to natural disasters by using call detail
  records to analyse the mobility of internally displaced persons.
\newblock {\em arXiv:1908.02381}, 2019.

\bibitem{Altuncu2019}
M.~Tarik Altuncu, Ayse~Seyyide Kaptaner, and Nur Sevencan.
\newblock Optimizing the access to healthcare services in dense refugee hosting
  urban areas: A case for istanbul, 2019.

\bibitem{Carmines1982}
Richard~A. Zeller, Steven~L. Nock, and Edward~G. Carmines.
\newblock Measurement in the social sciences: The link between theory and data.
\newblock {\em Contemporary Sociology}, 11(1):79, January 1982.

\bibitem{Marsden1984}
Peter~V. Marsden and Karen~E. Campbell.
\newblock Measuring tie strength.
\newblock {\em Social Forces}, 63(2):482, December 1984.

\bibitem{Marsden2012}
P.~V. Marsden and K.~E. Campbell.
\newblock Reflections on conceptualizing and measuring tie strength.
\newblock {\em Social Forces}, 91(1):17--23, August 2012.

\bibitem{Karsai2014}
M{\'{a}}rton Karsai, Nicola Perra, and Alessandro Vespignani.
\newblock Time varying networks and the weakness of strong ties.
\newblock {\em Scientific Reports}, 4(1), February 2014.

\bibitem{Krings2012}
Gautier Krings, M{\'{a}}rton Karsai, Sebastian Bernhardsson, Vincent~D Blondel,
  and Jari Saram\"{a}ki.
\newblock Effects of time window size and placement on the structure of an
  aggregated communication network.
\newblock {\em {EPJ} Data Science}, 1(1), May 2012.

\bibitem{Kovanen2013}
L.~Kovanen, K.~Kaski, J.~Kertesz, and J.~Saramaki.
\newblock Temporal motifs reveal homophily, gender-specific patterns, and group
  talk in call sequences.
\newblock {\em Proceedings of the National Academy of Sciences},
  110(45):18070--18075, October 2013.

\bibitem{Blondel2015}
Vincent~D Blondel, Adeline Decuyper, and Gautier Krings.
\newblock A survey of results on mobile phone datasets analysis.
\newblock {\em {EPJ} Data Science}, 4(1), August 2015.

\bibitem{Park2018}
Patrick~S. Park, Joshua~E. Blumenstock, and Michael~W. Macy.
\newblock The strength of long-range ties in population-scale social networks.
\newblock {\em Science}, 362(6421):1410--1413, December 2018.

\bibitem{TimeAllocation}
Giovanna Miritello, Rub{\'{e}}n Lara, and Esteban Moro.
\newblock Time allocation in social networks: Correlation between social
  structure and human communication dynamics.
\newblock In {\em Understanding Complex Systems}, pages 175--190. Springer
  Berlin Heidelberg, 2013.

\bibitem{Saramaki2014}
J.~Saramaki, E.~A. Leicht, E.~Lopez, S.~G.~B. Roberts, F.~Reed-Tsochas, and
  R.~I.~M. Dunbar.
\newblock Persistence of social signatures in human communication.
\newblock {\em Proceedings of the National Academy of Sciences},
  111(3):942--947, jan 2014.

\bibitem{Granovetter}
Mark~S. Granovetter.
\newblock The strength of weak ties.
\newblock {\em American Journal of Sociology}, 78(6):1360--1380, 1973.

\bibitem{Wuchty2009}
S.~Wuchty.
\newblock What is a social tie?
\newblock {\em Proceedings of the National Academy of Sciences},
  106(36):15099--15100, sep 2009.

\bibitem{Wellman1982}
B~Wellman.
\newblock Studying personal communities.
\newblock {\em Social Networks}, page e26972, 1990.

\bibitem{Granovetter1985}
Mark Granovetter.
\newblock Economic action and social structure: The problem of embeddedness.
\newblock {\em American Journal of Sociology}, 91(3):481--510, November 1985.

\bibitem{Friedkin1990}
Noah~E. Friedkin.
\newblock A guttman scale for the strength of an interpersonal tie.
\newblock {\em Social Networks}, 12(3):239--252, September 1990.

\bibitem{Brewer2000}
Devon~D. Brewer.
\newblock Forgetting in the recall-based elicitation of personal and social
  networks.
\newblock {\em Social Networks}, 22(1):29--43, May 2000.

\bibitem{Wilmot1987}
William Wilmot.
\newblock {\em Dyadic Communication}.
\newblock Random House, 1987.
\newblock ISBN: 0394358260.

\bibitem{Dindia1993}
Kathryn Dindia and Daniel~J. Canary.
\newblock Definitions and theoretical perspectives on maintaining
  relationships.
\newblock {\em Journal of Social and Personal Relationships}, 10(2):163--173,
  May 1993.

\bibitem{Hallinan1978}
Maureen~T Hallinan.
\newblock The process of friendship formation.
\newblock {\em Social Networks}, 1(2):193--210, January 1978.

\bibitem{Burt2000}
Ronald~S Burt.
\newblock Decay functions.
\newblock {\em Social Networks}, 22(1):1--28, May 2000.

\bibitem{Burt2002}
Ronald~S. Burt.
\newblock Bridge decay.
\newblock {\em Social Networks}, 24(4):333--363, October 2002.

\bibitem{Backstrom2014}
Lars Backstrom and Jon Kleinberg.
\newblock Romantic partnerships and the dispersion of social ties.
\newblock In {\em Proceedings of the 17th ACM conference on Computer supported
  cooperative work {\&} social computing - CSCW 14}, 2014.

\bibitem{Navarro2017}
Henry Navarro, Giovanna Miritello, Arturo Canales, and Esteban Moro.
\newblock Temporal patterns behind the strength of persistent ties.
\newblock {\em {EPJ} Data Science}, 6(1), dec 2017.

\bibitem{Ayres1983}
Joe Ayres.
\newblock Strategies to maintain relationships: Their identification and
  perceived usage.
\newblock {\em Communication Quarterly}, 31(1):62--67, January 1983.

\bibitem{Feld1981}
Scott~L. Feld.
\newblock The focused organization of social ties.
\newblock {\em American Journal of Sociology}, 86(5):1015--1035, March 1981.

\bibitem{Kahanda2009}
Indika Kahanda and Jennifer Neville.
\newblock Using transactional information to predict link strength in online
  social networks.
\newblock In {\em Third International AAAI Conference on Weblogs and Social
  Media}, 2009.

\bibitem{Miritello2013b}
Giovanna Miritello.
\newblock {\em Temporal Patterns of Communication in Social Networks}.
\newblock Springer International Publishing, 2013.

\bibitem{JariTemp}
Petter Holme and Jari Saram\"{a}ki.
\newblock Temporal networks.
\newblock {\em Physics Reports}, 519(3):97--125, oct 2012.

\bibitem{Wang2011}
Dashun Wang, Dino Pedreschi, Chaoming Song, Fosca Giannotti, and Albert-Laszlo
  Barabasi.
\newblock Human mobility, social ties, and link prediction.
\newblock In {\em Proceedings of the 17th ACM SIGKDD international conference
  on Knowledge discovery and data mining - KDD}, 2011.

\bibitem{Raeder2011}
Troy Raeder, Omar Lizardo, David Hachen, and Nitesh~V. Chawla.
\newblock Predictors of short-term decay of cell phone contacts in a large
  scale communication network.
\newblock {\em Social Networks}, 33(4):245--257, October 2011.

\bibitem{MiritelloStrategy}
Giovanna Miritello, Esteban Moro, Rub{\'{e}}n Lara, Roc{\'{\i}}o
  Mart{\'{\i}}nez-L{\'{o}}pez, John Belchamber, Sam~G.B. Roberts, and
  Robin~I.M. Dunbar.
\newblock Time as a limited resource: Communication strategy in mobile phone
  networks.
\newblock {\em Social Networks}, 35(1):89--95, jan 2013.

\bibitem{Wuchty2011}
Stefan Wuchty and Brian Uzzi.
\newblock Human communication dynamics in digital footsteps: A study of the
  agreement between self-reported ties and email networks.
\newblock {\em {PLoS} {ONE}}, 6(11):e26972, nov 2011.

\bibitem{Gilbert2009}
Eric Gilbert and Karrie Karahalios.
\newblock Predicting tie strength with social media.
\newblock In {\em Proceedings of the 27th international conference on Human
  factors in computing systems - {CHI} 09}, 2009.

\bibitem{KarsaiBursty}
M{\'{a}}rton Karsai, Kimmo Kaski, Albert-L{\'{a}}szl{\'{o}} Barab{\'{a}}si, and
  J{\'{a}}nos Kert{\'{e}}sz.
\newblock Universal features of correlated bursty behaviour.
\newblock {\em Scientific Reports}, 2(1), May 2012.

\bibitem{Goh2008}
K.-I. Goh and A.-L. Barab{\'{a}}si.
\newblock Burstiness and memory in complex systems.
\newblock {\em {EPL} (Europhysics Letters)}, 81(4):48002, jan 2008.

\bibitem{Barabsi2005}
Albert-L{\'{a}}szl{\'{o}} Barab{\'{a}}si.
\newblock The origin of bursts and heavy tails in human dynamics.
\newblock {\em Nature}, 435(7039):207--211, may 2005.

\bibitem{Heydari2018}
Sara Heydari, Sam~G. Roberts, Robin I.~M. Dunbar, and Jari Saram\"{a}ki.
\newblock Multichannel social signatures and persistent features of ego
  networks.
\newblock {\em Applied Network Science}, 3(1), May 2018.

\bibitem{Jo2019Burst}
Hang-Hyun Jo, Takayuki Hiraoka, and Mikko Kivel{\"a}.
\newblock Burst-tree decomposition of time series reveals the structure of
  temporal correlations.
\newblock {\em arXiv:1907.13556 [physics.data-an]}, 2019.

\bibitem{Kivela2015}
Mikko Kivel\"{a} and Mason~A. Porter.
\newblock Estimating interevent time distributions from finite observation
  periods in communication networks.
\newblock {\em Physical Review E}, 92(5), nov 2015.

\bibitem{Kivela2012}
Mikko Kivel\"{a}, Raj~Kumar Pan, Kimmo Kaski, J{\'{a}}nos Kert{\'{e}}sz, Jari
  Saram\"{a}ki, and M{\'{a}}rton Karsai.
\newblock Multiscale analysis of spreading in a large communication network.
\newblock {\em Journal of Statistical Mechanics: Theory and Experiment},
  2012(03):P03005, mar 2012.

\bibitem{Holme2015}
Petter Holme and Fredrik Liljeros.
\newblock Birth and death of links control disease spreading in empirical
  contact networks.
\newblock {\em Scientific Reports}, 4:4999, 2015.

\bibitem{Panda2002}
Satchidananda Panda, John~B. Hogenesch, and Steve~A. Kay.
\newblock Circadian rhythms from flies to human.
\newblock {\em Nature}, 417(6886), 2002.

\bibitem{Aledavood2017}
Talayeh Aledavood, Sune Lehmann, and Jari Saram\"{a}ki.
\newblock Social network differences of chronotypes identified from mobile
  phone data.
\newblock {\em {EPJ} Data Science}, 7(1), October 2018.

\bibitem{Wittmann2006Social}
Marc Wittmann, Jenny Dinich, Martha Merrow, and Till Roenneberg.
\newblock Social jetlag: Misalignment of biological and social time.
\newblock {\em Chronobiology International}, 23(1-2):497--509, January 2006.

\bibitem{aledavood2015a}
Talayeh Aledavood, Sune Lehmann, and Jari Saramäki.
\newblock Digital daily cycles of individuals.
\newblock {\em Frontiers in Physics}, 3:73, 2015.

\bibitem{Aledavood2015b}
Talayeh Aledavood, Eduardo L{\'{o}}pez, Sam G.~B. Roberts, Felix Reed-Tsochas,
  Esteban Moro, Robin I.~M. Dunbar, and Jari Saram\"{a}ki.
\newblock Daily rhythms in mobile telephone communication.
\newblock {\em {PLOS} {ONE}}, 10(9):e0138098, September 2015.

\bibitem{Aledavood2015c}
Talayeh Aledavood, Eduardo López, Sam G.~B. Roberts, Felix Reed-Tsochas,
  Esteban Moro, Robin I.~M. Dunbar, and Jari Saramäki.
\newblock Channel-specific daily patterns in mobile phone communication, 2015.

\bibitem{Adan2012}
Ana Adan, Simon~N. Archer, Maria~Paz Hidalgo, Lee~Di Milia, Vincenzo Natale,
  and Christoph Randler.
\newblock Circadian typology: A comprehensive review.
\newblock {\em Chronobiology International}, 29(9), 2012.

\bibitem{Matthews1975}
B.W. Matthews.
\newblock Comparison of the predicted and observed secondary structure of t4
  phage lysozyme.
\newblock {\em Biochimica et Biophysica Acta ({BBA}) - Protein Structure},
  405(2):442--451, October 1975.

\bibitem{Boughorbel2017}
Sabri Boughorbel, Fethi Jarray, and Mohammed El-Anbari.
\newblock Optimal classifier for imbalanced data using matthews correlation
  coefficient metric.
\newblock {\em {PLOS} {ONE}}, 12(6):e0177678, June 2017.

\bibitem{Louppe2013}
Gilles Louppe, Louis Wehenkel, Antonio Sutera, and Pierre Geurts.
\newblock Understanding variable importances in forests of randomized trees.
\newblock In C.~J.~C. Burges, L.~Bottou, M.~Welling, Z.~Ghahramani, and K.~Q.
  Weinberger, editors, {\em Advances in Neural Information Processing Systems
  26}, pages 431--439. Curran Associates, Inc., 2013.

\end{thebibliography}



\section*{Supplementary material}

# S1 Analysis of Variables

## S1.1 Static and Dynamic Overlap

In this paper, we find proxies for measuring the strength of ties in communication networks by claiming that tie strength manifests both in communication patterns and network topology. We measure embeddedness via average overlap over 1-month aggregation windows. More explicitly, the overlap obtained over a shorter time period $\Delta T$ and using a sliding window approach as to obtain a time series of overlap values $\{O_i\}$. In our paper, we use a $\Delta T = 1$ month as an aggregation period, and here we compare this average dynamic measure to overlap measured on the full aggregation window.

Figure 1 displays the joint distribution of static and dynamic overlap measures. Both network statistics have similar distributions, with Pearson's correlation coefficient of 0.77 and a rank correlation of 0.81. The main difference between is the larger number of zero-valued entries for dynamic overlap, up to 5.18% as opposed to static overlap's 1.03%. We briefly see how the set of common neighbors ($\mathcal{N}_i \cup \mathcal{N}_j$) changes from static to dynamic overlap. We divide the set of common neighbors onto three cases depending on their stability across the dynamic overlap's aggregation period. We find that only 1.32% of neighbors were connected to both nodes in a tie at all aggregation windows, 50.66% neighbors were connected to both nodes during at least one aggregation window and 48.02% of neighbors not connected to both nodes during the same aggregation period. This high turnover of common neighbors helps explain why dynamic overlap tends to have smaller values than it's static counterpart, since at least 48% of neighbors from $O$ do not translate to common neighbors on $\hat{O}^t$.

## S1.2 Creation of weekly cluster profiles

Our analysis of weekly activity profiles involves a two-step procedure where we first have a high-granularity approach by dividing the week into $n = 168$ hours, followed by a clustering process to reduce the number of variables. There is a trade-off between the number of variables and the information they contain on overlap—a large number of variables implies that for most ties, their weekly profiles will be zero-valued, whereas having a low number of variables (e.g., two weekly profile variables of calls placed on work and leisure times) might ignore valuable details on which topological information is encoded at certain times.

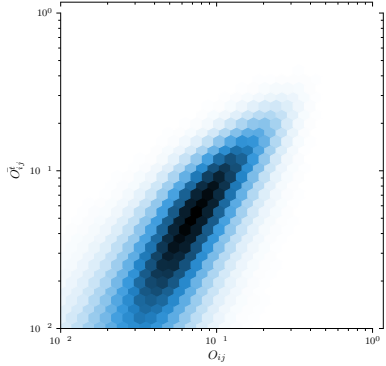

**Figure S1.** Joint distribution of static $O$ and dynamic $\hat{O}^t$ overlap. For visualization purposes we only include positive overlap values.

To determine our clusters, we selected a sample of 100,000 ties and for each one obtained 168 weekly activity variables $\{\phi^h\}_{h=1\ldots168}$, where $\phi^h$ contains the fraction of calls placed at hour $h$. We then computed the correlation matrix to detect timing where behaviour is similar. We used Markov Cluster Algorithm (MCA) [1], a method that uses the correlation matrix as input, as well as a parameter $\psi$ that determines a cutoff value that determines the granularity of the clusters (which we denote by $C_\psi$). Given $\psi$, we can determine the weekly clusterized profiles $\{\phi^c\}_{c \in C_\psi}$, and we then determine the smallest cutoff value that captures as much diversity in overlap values as our high-granularity approach, our criteria for clusterization.

## S1.3 Feature correlations

In this paper we use features from existing modelling approaches to communication found in the literature, along with new ones we introduce. In the main body of the text we describe how some of these variables are theoretically related to each other, but we do not provide an empirical analysis of our own dataset. On Figure 2 we depict the correlation between our features sorted by modelling approach: intensity, inter-event time distribution, temporal stability, bursty cascades, distribution of busty cascades, and weekly behaviour signatures. There is a high degree of correlation between different features of communication, both within groups and between groups, where the clusters of weekly activity patterns encompass the lowest correlations. This is expected, as we derived them using a methodology that creates clusters based on correlated behaviour.

## S1.4 Relationship to overlap correcting for $w$

We present plots Figures 3, 4, 5 that depict relationships of the type $\langle O|F \rangle$, for all features $F$. Since one of the main goals of this research is to understand whether the features provide additional information not captured by communication intensity $w$; we correct for part of the effect of $w$ by plotting $\langle O|F \rangle$ at five different levels of $w$, determined by equal-sized quantiles.

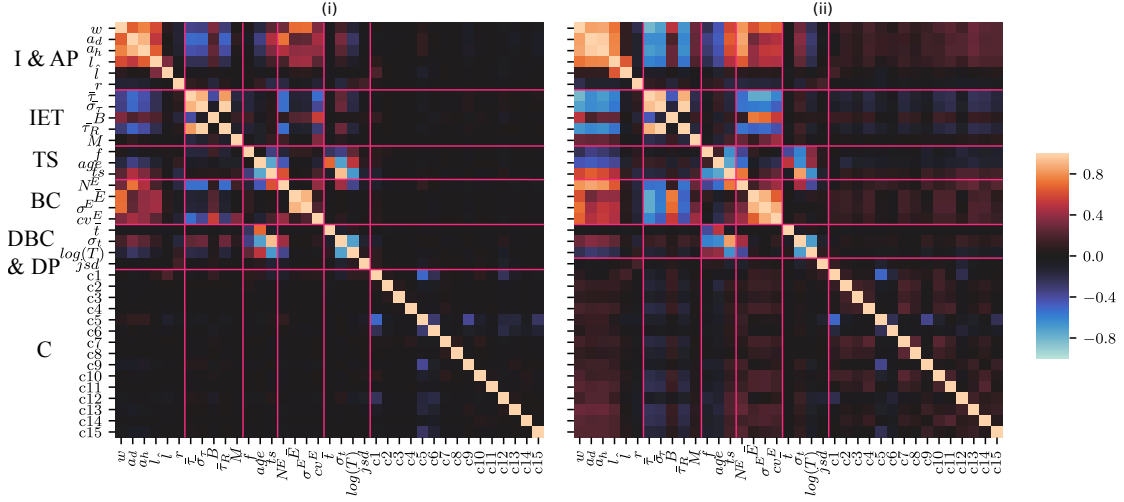

**Figure S2.** Feature correlation matrices measured by (i) Pearson's and (ii) Spearman's correlation coefficients. Features are sorted according to their modelling approach, each case divided by pink lines. Features display high withing-group correlations, with lower between group correlations. Notably, intensity and static features (first group) have a mostly negative correlation to IET-derived features (second group) and a positive correlation to features derived from bursty cascades (third group). Weekly clusters (last group) show no relevant correlations among themselves or to other variables, with two main exceptions: negative correlations between clusters $C1$ (late night) and $C5$ (weekday worktimes), and clusters $C5$ and $C9$ (weekend night).

We observe that for a large set of features an increase in $F$ implies changes in both the average overlap and the overlap distribution. These figures provide evidence that for most features the relationship to overlap is not necessarily linear. Indeed, our features seem to both interact with communication intensity $w$ and encode different information about overlaps at different values. For instance, for the number of days and hours with contacts ($a_d$ and $a_h$ on Figure 3), the average overlap increases almost linearly with $F = a_d, a_h$ for communication intensity $w \leq 25$; overlap increases at a decreasing rate with $F = a_d, a_h$ for higher communication intensity. Now, considering the average IET ($\bar{\tau}$ on Figure 3), we find overlap to be more sensitive to changes in $\bar{\tau}$ when it is smaller, and after a certain degree $\bar{\tau}$ becomes less effective at encoding information of overlap. Most relationships seems to be both non-trivial and non-linear, we expect ML models that favour both variable interactions and non-linearities to be more efficient at capturing overlap.

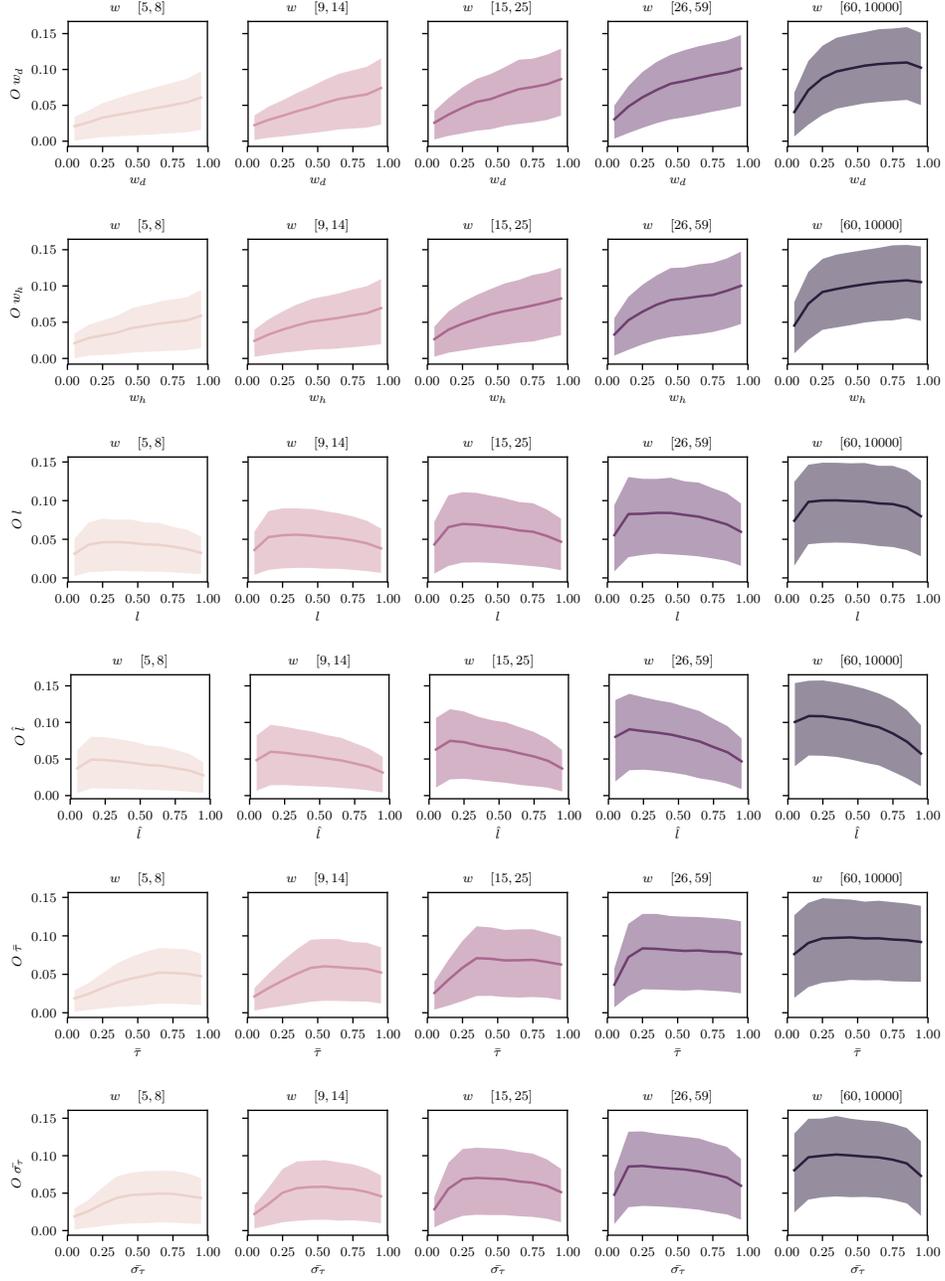

**Figure S3.** Average topological overlap for non-cluster variables correcting for different levels of communication intensity ($w$, in columns). Shade includes 80% of the distribution. From top to bottom: number of days with contacts ($a_d$), number of hours with contacts ($a_h$), total call length or duration ($l$), average call length ($\hat{l}$), average IET ($\bar{\tau}$), standard deviation of the IET ($\bar{\sigma}_\tau$).

4

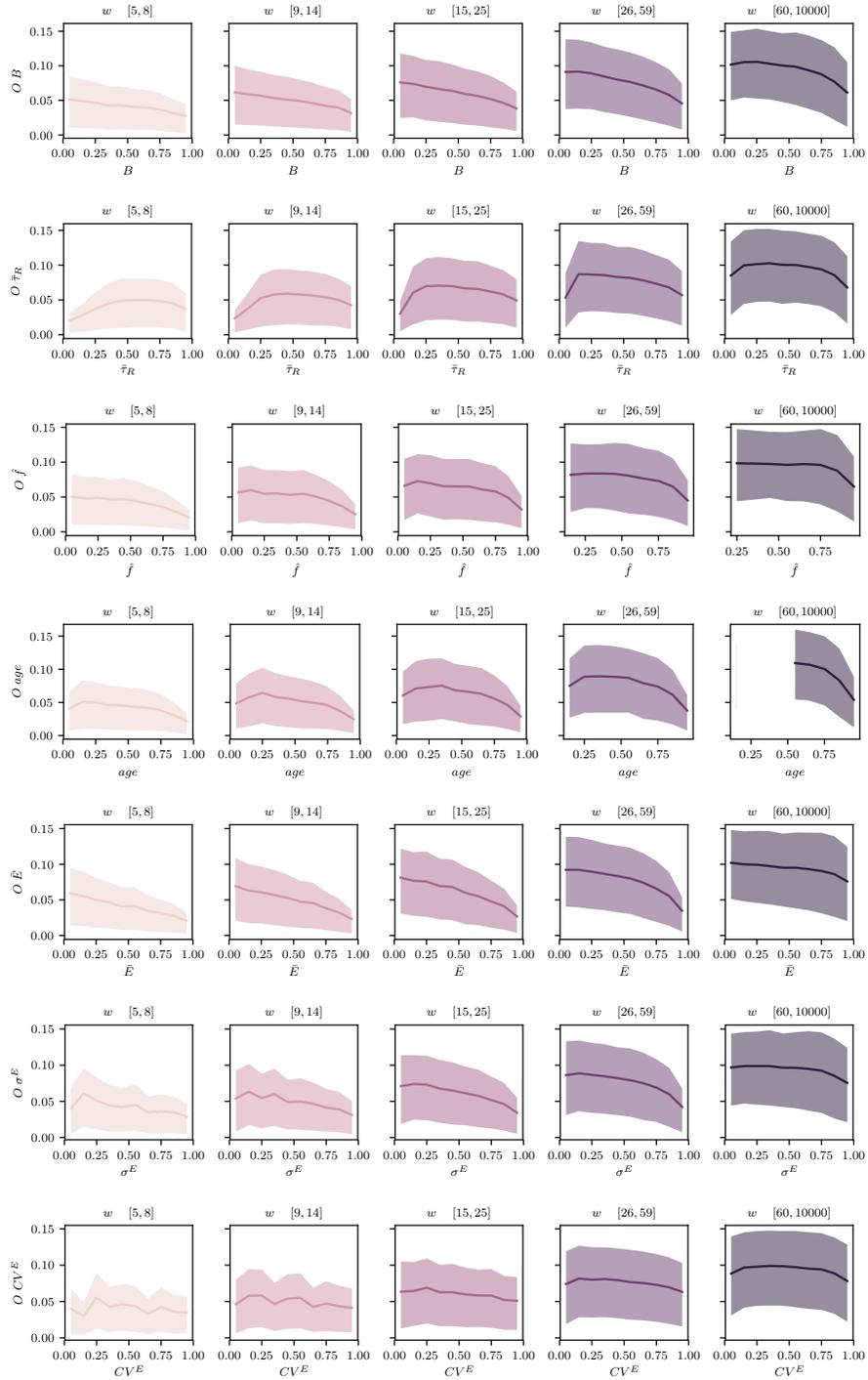

**Figure S4.** Average topological overlap for non-cluster variables correcting for different levels of communication intensity ($w$, in columns). Shade includes 80% of the distribution. From top to bottom: burstiness ($B$), average inter-relay time ($\bar{\tau}_R$), relative freshness ($\hat{f}$), tie age ($age$), average number of calls per bursty train ($\bar{E}$), std. deviation of number of calls per bursty train ($\sigma^E$), coefficient of variation of $E$ ($CV^E$).

5

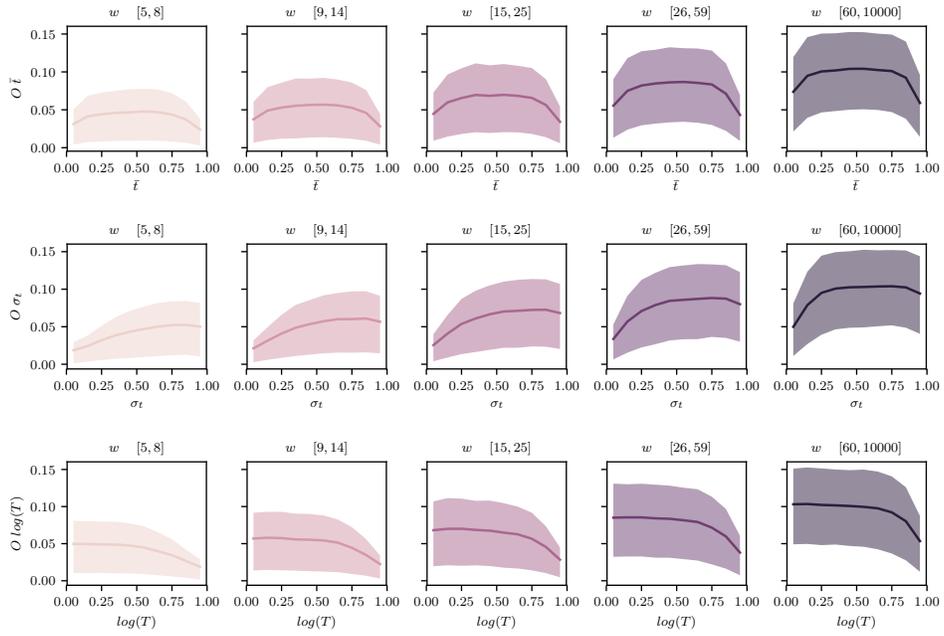

**Figure S5.** Average topological overlap for non-cluster variables correcting for different levels of communication intensity ($w$). From top to bottom:

# S2 Predicting Static Overlap

The main body of the text focused on predicting a dynamic measure of overlap on one-month aggregation windows, $\hat{O}^t$. Here we present results where the predictive variable is overlap for the full four-month aggregation window, which we refer to as static overlap $O$. The conditions of the experiment were identical to those with $\hat{O}^t$: we used the same random sample of 500,000 ties, and predicted different high/low overlap values. We used 3-fold cross validation with four machine learning models ABC, RF, LC and QDA in three scenarios: using each feature as a predictor; using $(F, X)$ variable combinations as predictors, where $F = a_d, a_h, N^E, w$ are the three best-performing variables along with communication intensity, and $X$ are all the features; and using all the features (the static overlap MCC scores were already presented in Figure 6 of the main text to serve as a point of reference).

Figure 6 our results of static overlap prediction in the single and dual-variable scenarios. The results are relatively consistent to the analysis of overlap prediction in the dynamic scenario, where the number of active days and hours, and the number of bursty trains are the best-performing variables. Although specific variable rankings differ, most variable's behaviour relative to $\alpha$ seem to be consistent (for instance, $C^*$ covers a large range of values, while $\sigma_t$'s performance is skewed towards lower values). The average performance of dual-variable models differs, however, with the single combination $(F, JSD)$ notably dominating all the other interactions, with a large difference in average performance with the following two combinations, $(F, \hat{l})$ and $(F, l)$, the average and total call length. Compared to dynamic overlap, there is much less variability in combinations $(F, X)$, both regarding $F$ (given a feature X on the x-axis, variables perform similarly) and regarding $X$ (given a top-performing feature $F$, most variables $X$ perform similarly, with the exception of $X = JSD$).

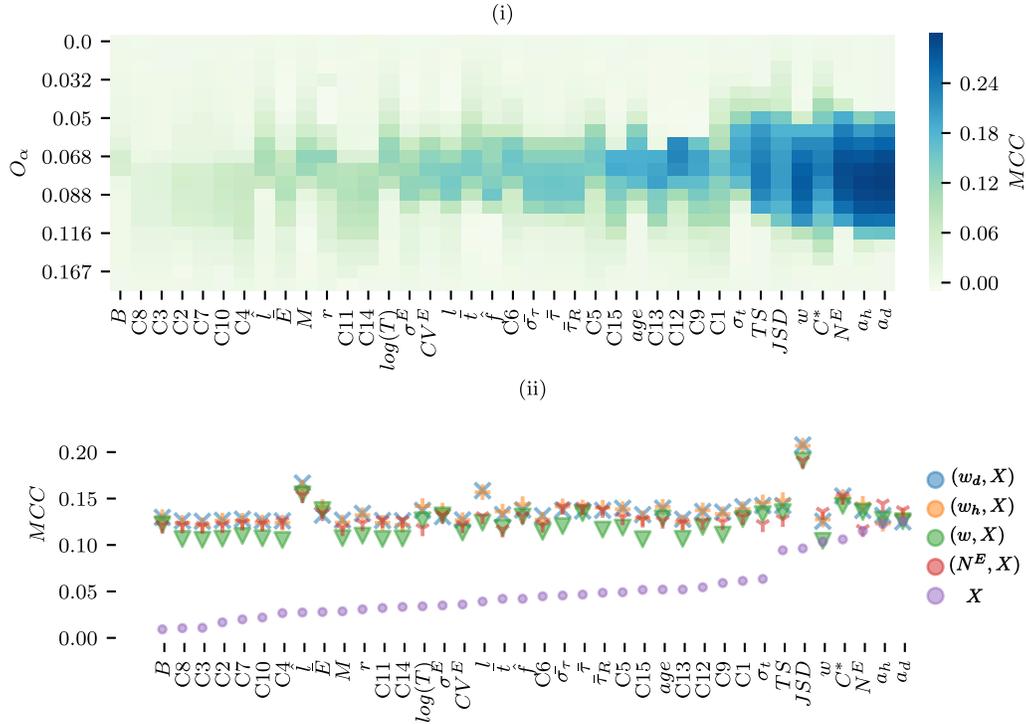

**Figure S6.** Matthew's Correlation Coefficient (MCC) for static overlap prediction, where the $x$-axis represents variables, the $y$-axis represents different cutoff values $\alpha$ for binary classification of high/low overlap, and the color represents MCC. Variables are ranked by their average performance over all cutoff values $\alpha$. ($i$) Average MCC for four models trained with single-feature predictors, where each variable is used to predict static overlap using RF, ABC, LG and QDA ($ii$) Comparison between single and dual-variable models, where we depict the average across all models.

# S3 Results by Machine Learning Model

We used four ML models to predict embeddedness in two different scenarios (static and mean temporal overlap) using temporal features of human communication, and present the average behaviour of the four models to rank variable performance. Feature performance, however, varied according to the ML models and overlap cutoff values. This can be expected as different ML models focus on different data aspects and might assume different data distributions. In this section we discuss how models affected variable rankings and overall performance.

## S3.1 Static Overlap

Figure 7 displays the feature performance in predicting static overlap. Feature importance for the highest-ranked variables is consistent across models, with temporally-aggregated intensity features $a_d$ and $a_h$ and $N^E$ dominating all rankings, followed in most cases by the aggregate number of calls $w$. In all cases the range of $O_\alpha$ values with discriminative capacity is similar across models. Our variable for differences in daily behaviour $JSD$ ranks highly for most models except for RF, where it has a similarly poor performance across all $O_\alpha$ values. In the best performing model, ABC, $C^*$ has the highest ranking, which could imply that a characterization of weekly behaviour contains the most information of tie strength, albeit in a possibly non-linear manner. For this model, $JSD$ has a large range of predictive capacity. Temporal stability $TS$ also displays a consistently high-ranking, with MCC values mostly on par with $w$.

The predictive capacity of individual cluster values varies highly depending on the models. Notably, $C12$ has the overall highest MCC values for clusters - yet its predictive range is limited, yielding a relatively poor average performance. Cluster features tend to rank higher in models that do not assume linearity in the data, such as RF and ABC. In these cases, clusters $C9$, $C12$, $C13$ and $C15$, all associated to nighttime or weekends, have a higher performance. Notably, the cluster associated to late nights, $C1$, ranks highly as it covers a wide range of values with lower predictive performance.

## S3.2 Dynamic Overlap

Measuring neighborhood overlap in a dynamic manner yields some key differences in our results when compared to the static version. While $a_d$ remains the highest-ranked feature, there is a higher feature turnover between $a_h$, $N^E$ and $TS$, with $w$ now ranked between the fifth and eight position depending on the model. Variables that were not highly ranked for the static scenario are now more prominent, such as $age$ and $\sigma_t$, which occurs consistently in all four models. The considerably large range of predictive values of $C^*$ in ABC and RF is now diminished and mostly on par with $a_h$, $a_d$ and $N^E$.

Our feature of differences in daily behaviour $JSD$ also ranks consistently lower in this scenario. As with the static case, it performs poorly with RF. For weekly signature clusters, we find a higher prominence in non-linear models RF and ABC. With mean temporal overlap, however, feature $C12$ is now the highest-ranked cluster in all but one model, having a similar behaviour of a small range with more intense predictive value.

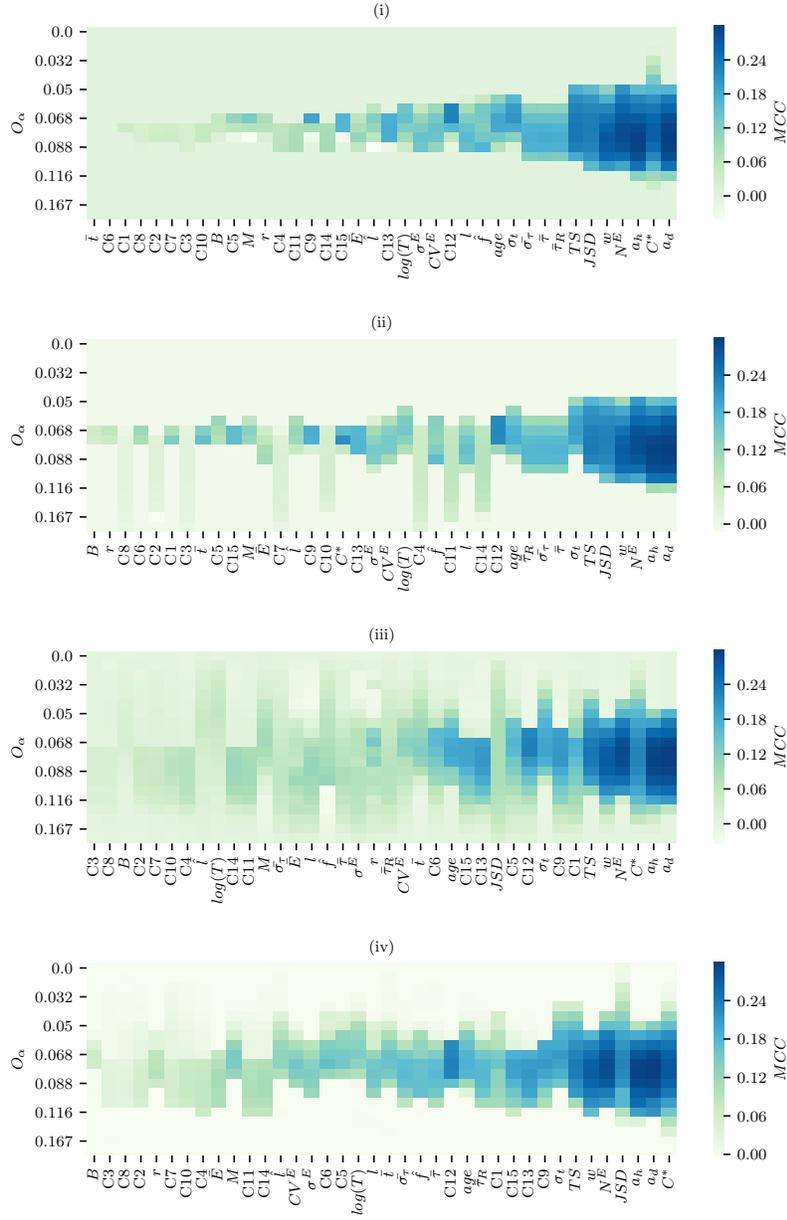

**Figure S7.** Feature performance in static overlap prediction for four different ML models ; (i) LR, (ii) QDA, (iii) RF and (iv) ABC. X-axes represent features ranked by average predictive performance, y-axes represent overlap cutuff values $O_\alpha$ and color represents model performance as measured by MCC.

**Figure S8.** Feature performance in dynamic overlap prediction for four different ML models ; (i) LR, (ii) QDA, (iii) RF and (iv) ABC. X-axes represent features ranked by average predictive performance, y-axes represent overlap cutuff values $O_\alpha$ and color represents model performance as measured by MCC.

## S4 Analysis of bursty cascades

We analyze the effect of the $\Delta t$ parameter on bursty trains and its relationship to overlap. In order to estimate the effect, we use a sample of 100,000 ties, and use vary $\Delta t$ on a grid of values with one-hour increments. We evaluate performance via (i) the MCC of a LR that classifies weak/strong ties based on overlap defined at $\alpha = 0.08$, and (ii) Pearson's correlation coefficient with overlap, with results depicted on Figure 9. We choose LR since it is computationally efficient when runnng a large number of cases, and since we know it to perform well for some many of the variables of interest.

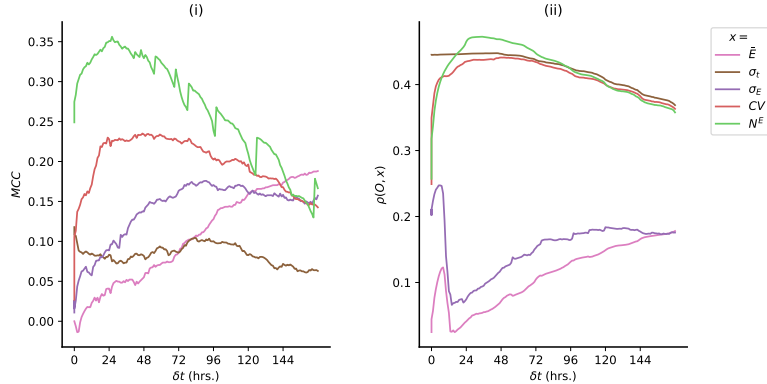

**Figure S9.** Effect of $\Delta t$ on the relationship between overlap and variables derived from bursty trains ($x$). (i) MCC score for LR model for high/low overlap defined at $\alpha = 0.08$ (ii) Pearson correlation coefficient between variables $x$ and overlap.

Our results suggest some variation both in predictive performance and correlation for different $\Delta t$ values and for different variables, yet this variation occurs slowly. Prominently, $N^E$ or the number of bursty trains has a predictive performance and correlation that peak at $\Delta t = 26$ and $\Delta t = 32$ hours respectively. We note two particular effects: the dependence of $\Delta t$ on daily 24-hour cycles, and the interplay between decreasing and increasing trends for different types of variables. In the first case, we note that $N^E$ varies on 24-hour cycles, which is likely due to how $\Delta t$ overriding the effect of daily call-placement cycles. Second, $N^E$ and $CV$ display an initial increase followed by a slower decrease in performance, which constrats to the slower increasing trends of $\bar{E}$ and $\sigma_E$. Indeed, since the number of calls $w$ remains constant, for large $\Delta t$ values the number of cascades decreases, while the number of calls in a cascade $E$ increases, in a manner transferring the information from one variable to the other.

### S4.1 Variable Correlations

Figure 10 displays the correlations between the $N^E$ defined for different $\Delta t$ values. Roughly, the two matrices display high positive correlation, particularly around the diagonal, yet we find three key aspects worth discussing, mainly related to Pearson's correlation matrix.

First, $N^E$ is more sensitive to smaller $\Delta t$, where $\Delta t \leq 1$ hour and $\Delta t \leq 1$ day characterize values that correlate the least to the rest of the matrix. That is, the higher-resolution $\Delta t$ values corrects for the most bursty behaviour. Indeed, since most links have a relatively small number of calls (with higher IET times), the number of bursty trains changes at a slower rate. Second, the matrix roughly follows a block

12

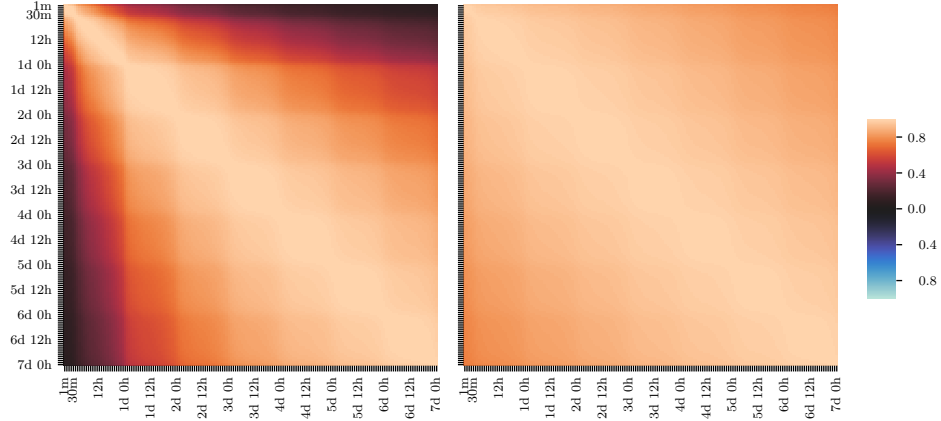

**Figure S10.** Correlations for number of bursty trains defined obtained for different $\Delta t$ values; (*left*) Pearson's and (*right*) Spearman's correlation coefficients. We explore $\Delta t$ values at $\Delta t = 1, 3, 5, 10, 15, 30, 45\,60$ minutes initially, followed by 1 hour increases.

structure, with high correlation on blocks around the diagonal, but where roughly 1-day blocks determine higher or lower correlation, which means that circadian patterns determine the correlation blocks. Last, these differences are decidedly less pronounced for Spearman's correlation coefficient, which implies that the ranking generated by $N^E$ at different $\Delta t$ values is roughly the same.

## S4.2 Number of Bursty Trains

We examine the effect of $\Delta t$ on the predictive capacity of $N^E$ for different cutoff values of both static and dynamic overlap. The predictive capacity of $N^E$ is greater at around $\Delta t = 1$ day, and does not vary greatly on small changes of $\Delta t$. The differences in performance seem to be greater for large differences in $\Delta t$.

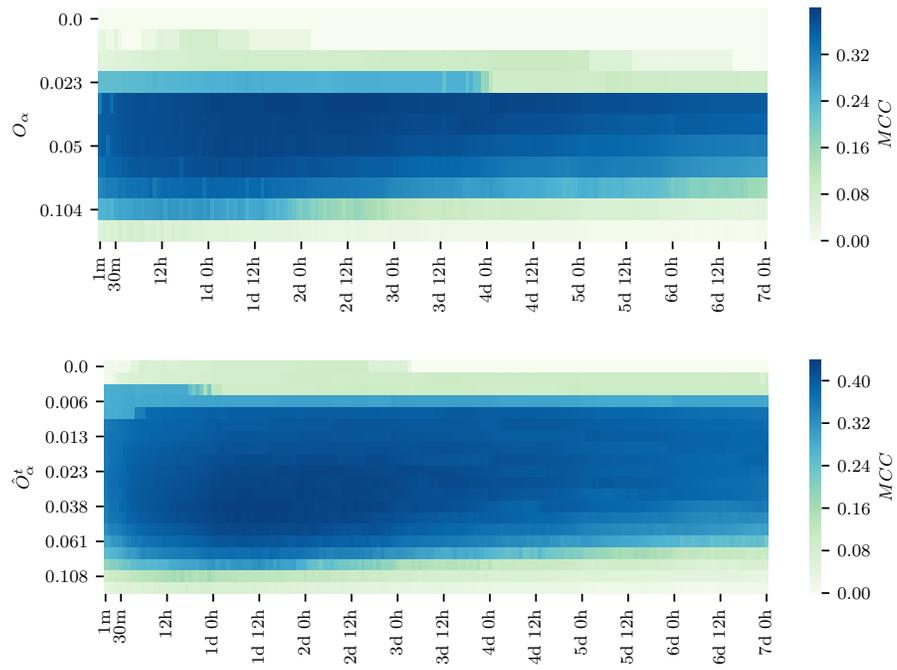

**Figure S11.** Effect of $\Delta t$ for using $N^E$ in the prediction of (*top*) static overlap and (*bottom*) mean temporal overlap. The x-axes represent $\Delta t$ values used to obtain $N^E$, the y-axes represent static overlap cutoff values $O_\alpha$, and color represents MCC

# S5   Weekly Signatures for Overlap Prediction

In this section, we examine the use of our clustered weekly signatures for overlap prediction. We fit predictive ML models to static and dynamic overlap using only the set of cluster features. Except for the input features, the conditions are the same as in the main paper: we predict binary low/high overlap for different cutoff values, using four ML models (RF, ABC, LR, QDA). We use a random sample of 500,000 ties and 3-fold cross validation to obtain scores and feature importance for the models where it is available.

Figure 12 depicts the MCC scores for each model at each cutoff value, as well as the feature importance of the best performing models (ABC for both cases). These weekly signature models achieve a performance similar to the best-performing variables in the other scenarios. Notably, there are strong performance differences per model, where LR yields an overall poor predictive capacity with a large drop when $\alpha$ is larger. Both ABC and RF capture higher-degree nonlinearities, which together with the poor LR performance could point towards non-linear relationships in the weekly clusters. As with the previous case, our features are able to predict dynamic overlap with a slightly better capacity than the static case (MCC of 0.31 and 0.276, respectively).

Feature importance points towards clusters previously identified as being predictive of overlap for our dataset. The most relevant clusters are $C1$, $C5$ and $C12$, which roughly correspond to late night, weekday worktimes and weekend afternoon, respectively, a result in line with the individual feature performance from before. Our results suggest a nontrivial relationship between the timings of communication events and network topology.

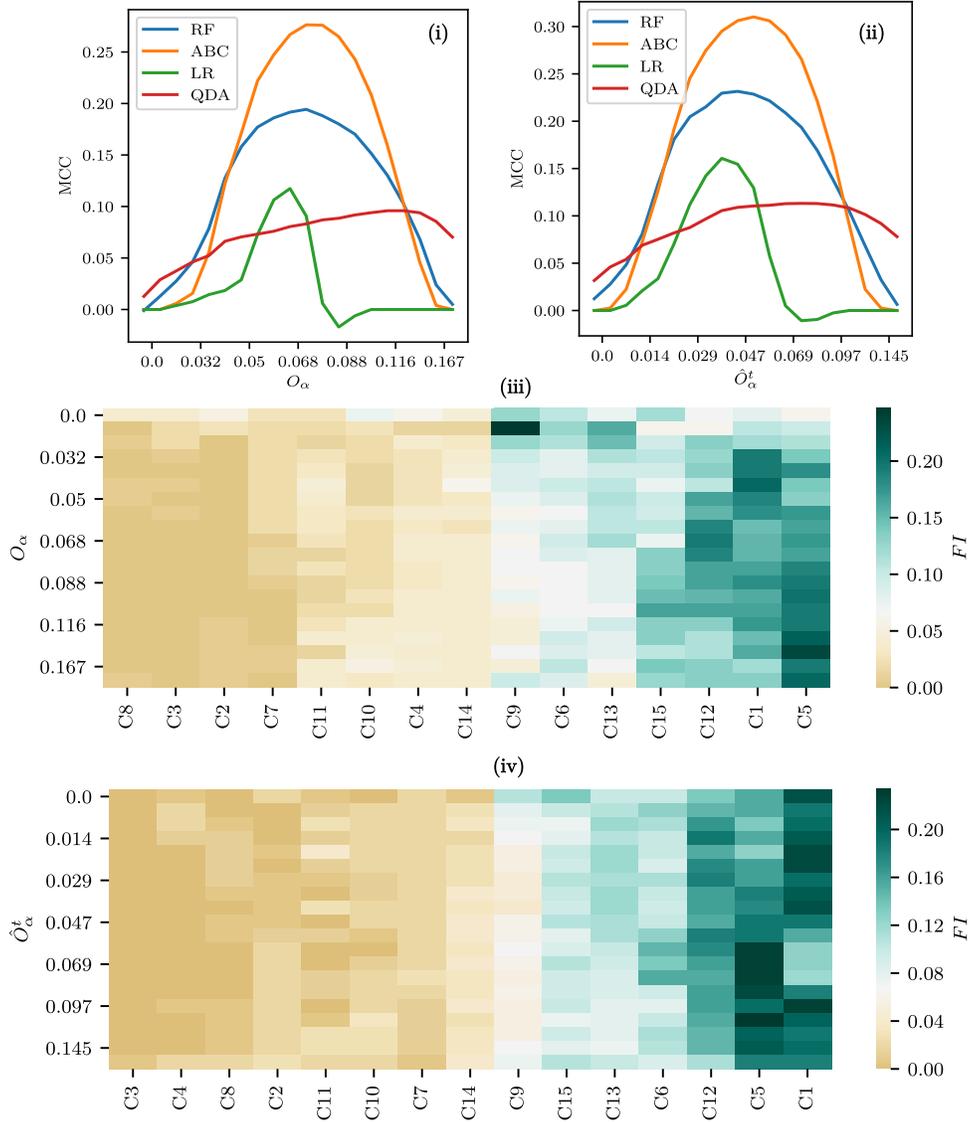

**Figure S12.** Full model scores and feature importance for weekly signatures. (*top*) MCC for four different models used in prediction of (*i*) static and (*ii*) dynamic overlap. (*iii - iv*) Feature importance (FI) for the overall best performing model, ABC, for prediction of (*iii*) static and (*iv*) dynamic overlap. Features are ranked by their average performance over all cutoff values $\alpha$.

# References

[1] Van Dongen, S.: A new cluster algorithm for graphs. Inf. Syst.1 (2002). doi:10.1046/j.1365-2575.2000.010001001.x17



\end{document}